\documentclass[11pt]{article}

\IfFileExists{english.ldf}{\usepackage[english]{babel}}{}
\usepackage[letterpaper,margin=1in]{geometry}

\usepackage{amsmath,amssymb,amsthm,amsfonts,bm}

\newtheorem{theorem}{Theorem}

\usepackage{graphicx}
\usepackage{booktabs}
\usepackage{natbib}
\usepackage[colorlinks=true,allcolors=blue]{hyperref}

\title{Deterministic Uniformization for Estimation and Latent Trajectory Inference in 
Interval-Censored Multistate Transition Models}
\author{
Yuxi Zhu\textsuperscript{1,2,*}
\\[1em]
\small
\textsuperscript{1}Department of Pediatrics, Rainbow Babies \& Children's Hospital, Cleveland, OH, USA
\\
\small
\textsuperscript{2}School of Medicine, Case Western Reserve University, Cleveland, OH, USA
\\[0.5em]
\small
\textsuperscript{*}Corresponding author: Yuxi.Zhu@uhhospitals.org
}
\date{}

\begin{document}
\maketitle

\begin{abstract}
Interval-censored multistate data arise when subjects are observed only at irregular visit times, 
leaving transition times and intermediate states unobserved. 
Likelihood-based inference for continuous-time Markov models typically requires repeated matrix exponential evaluations, 
which can become computationally burdensome in models with complex transition structures. 
We propose a deterministic uniformization expectation--maximization (EM) algorithm for estimation and 
posterior latent trajectory inference in interval-censored multistate transition models. 
The method represents the continuous-time process through an embedded discrete-time Markov chain indexed 
by latent Poisson event counts, yielding deterministic E-step calculations that avoid repeated matrix exponential evaluations 
within the EM iterations. The M-step reduces to transition-specific Poisson regressions with exposure offsets, 
allowing flexible covariate modeling of transition intensities. 
We establish uniform truncation error bounds for the deterministic uniformization approximation and 
show that the resulting procedure is an asymptotically monotone generalized EM algorithm. Under standard regularity and identifiability conditions, the resulting estimator is consistent and asymptotically normal. 
Simulation studies demonstrate accurate estimation and favorable computational performance relative to direct matrix-exponential 
likelihood optimization, particularly in models with complex latent trajectory structures. We further propose an exact dynamic-programming method 
for computing candidate pathway posterior probabilities and identifying the most probable latent clinical pathway between consecutive observations. 
Posterior expected transition counts and state occupancy times are also available as additional summaries,
while expected-count ratios provide a computational shortcut for pathway classification in simple mutually exclusive pathway
structures. These capabilities support both population-level 
inference and subject-specific characterization of disease progression.

\end{abstract}

\section{Introduction}

Multistate transition models are widely used to characterize disease progression, treatment response, behavioral development, and other longitudinal processes in which individuals move among a finite set of clinically or scientifically meaningful states \citep{kalbfleisch1985analysis}. Applications include diabetes progression \citep{zhu2025joint}, statin-induced adverse effects \citep{zhu2021multistate}, and many other biomedical settings where the underlying disease process evolves over time. These studies commonly involve discrete and irregular follow-up visits, leading to interval-censored multistate data. Consequently, states are observed only at visit times, while the exact transition times and intermediate states between visits remain unobserved.

Continuous-time Markov chains (CTMCs) provide a natural framework for modeling interval-censored multistate data. Under this framework, likelihood-based inference is typically based on transition probability matrices obtained from matrix exponentials of the generator matrix \citep{jackson2011multi}. Although conceptually straightforward, repeated matrix exponential evaluations during numerical optimization can become computationally demanding when the number of states, allowable transitions, or covariate-dependent intensity parameters increases \citep{williams2020bayesian, zhu2024uniformization}. Furthermore, direct likelihood optimization focuses primarily on parameter estimation and does not naturally provide posterior summaries of the latent trajectories connecting observed states.

To address these challenges,we propose a deterministic uniformization EM algorithm for interval-censored multistate data. Uniformization represents the CTMC transition probability matrix as a Poisson-weighted sum over powers of an embedded discrete-time transition matrix. This representation transforms the latent continuous-time process into a discrete sequence indexed by a latent Poisson event count, allowing expected transition counts and state occupancy times to be evaluated deterministically. As a result, the E-step avoids Monte Carlo sampling and repeated matrix exponential evaluations, while the M-step reduces to a collection of transition-specific Poisson regression problems.

The proposed framework offers three main advantages. First, it provides an efficient computational alternative to direct matrix-exponential likelihood optimization through recursive probability propagation under uniformization. Second, the resulting M-step admits flexible regression modeling of transition intensities and separates naturally across transitions, facilitating estimation and interpretation of covariate effects. Third, it enables subject-specific reconstruction of latent clinical pathways between observation times through an exact dynamic-programming calculation that sums over all latent uniformization sequences corresponding to each candidate clinical pathway. Posterior expected transition counts and state occupancy times remain available as additional summaries, and expected-count ratios provide a computational shortcut for pathway classification in simple mutually exclusive pathway structures. These features provide a foundation for both population-level inference and individualized characterization of disease progression, with potential applications to precision medicine and dynamic treatment strategies.

The remainder of the paper is organized as follows. Section~\ref{sec:model} introduces the interval-censored multistate transition model. Section~\ref{sec:method} presents the deterministic uniformization EM algorithm. Section~\ref{sec:theory} summarizes key theoretical properties, with full proofs and standard asymptotic results deferred to the Supplementary Material. Section~\ref{sec:simulation} describes the simulation studies, Section~\ref{sec:application} presents the real-data application, and Section~\ref{sec:discussion} concludes.

\section{Model Setup}\label{sec:model}

\subsection{Continuous-time multistate transition model}

Let $X_i(t)$ denote the state occupied by subject $i$ at time $t$, with finite state space $\mathcal S=\{1,\ldots,S\}$. Conditional on subject-level covariates $x_i$, we assume that $X_i(t)$ follows a time-homogeneous CTMC with generator matrix $Q(x_i;\theta)$.

For each allowed transition $r\to s$, $r\ne s$, the transition intensity is modeled as
\begin{equation}
q_{rs}(x_i;\theta)=\exp(x_i^\top\beta_{rs}),
\label{eq:intensity}
\end{equation}
where $x_i$ includes an intercept and covariates, and $\beta_{rs}$ is a transition-specific coefficient vector. If transition $r\to s$ is not allowed, then $q_{rs}(x_i;\theta)=0$. The diagonal entries satisfy
\begin{equation}
q_{rr}(x_i;\theta)=-\sum_{s\ne r}q_{rs}(x_i;\theta).
\end{equation}

The transition probability matrix over an interval of length $t$ is
\begin{equation}
P(t\mid x_i;\theta)=\exp\{Q(x_i;\theta)t\}.
\label{eq:matrix_exp}
\end{equation}

\subsection{Panel observation structure}

Subject $i$ is observed at visit times $0\le t_{i0}<t_{i1}<\cdots<t_{im_i}$. The observed data are
\begin{equation}
\mathcal O_i=\{(t_{ij},Y_{ij}):j=0,\ldots,m_i\},
\end{equation}
where $Y_{ij}=X_i(t_{ij})$. For interval $(t_{i,j-1},t_{ij}]$, let
\[
a=Y_{i,j-1},\qquad b=Y_{ij},\qquad \Delta t_{ij}=t_{ij}-t_{i,j-1}.
\]
The observed interval likelihood contribution is
\begin{equation}
P_{ab}(\Delta t_{ij}\mid x_i;\theta)=
\left[\exp\{Q(x_i;\theta)\Delta t_{ij}\}\right]_{ab}.
\label{eq:interval_lik}
\end{equation}
Thus the observed-data log-likelihood is
\begin{equation}
\ell_n(\theta)=\sum_{i=1}^n\sum_{j=1}^{m_i}
\log P_{Y_{i,j-1},Y_{ij}}(\Delta t_{ij}\mid x_i;\theta).
\end{equation}

\section{Deterministic Uniformization EM Algorithm}\label{sec:method}

\subsection{Uniformization representation}

For subject \(i\),  choose a subject-specific dominating
uniformization rate
\[
\lambda_i
\ge
\sup_{\theta\in\Theta}
\max_r\{-q_{rr}(x_i;\theta)\}.
\]

Thus, \(\lambda_i\) is independent of \(\theta\) and dominates all state
exit rates uniformly over \(\Theta\).

and define the embedded transition matrix
\begin{equation}
R_i(\theta)=I+Q(x_i;\theta)/\lambda_i.
\end{equation}
Then $R_i(\theta)$ is stochastic and
\begin{equation}
\exp\{Q(x_i;\theta)t\}
=\sum_{k=0}^\infty e^{-\lambda_i t}\frac{(\lambda_i t)^k}{k!}R_i(\theta)^k.
\label{eq:uniformization}
\end{equation}
This representation interprets the CTMC as an embedded discrete-time Markov chain evaluated at latent Poisson event times. Conditional on interval length $t$, the number of embedded events satisfies
\begin{equation}
K\sim \mathrm{Poisson}(\lambda_i t).
\end{equation}

\subsection{Complete-data likelihood}

If the full latent path were observed, let $N_{rs,i}$ denote the number of transitions from state $r$ to state $s$ for subject $i$, and let $T_{r,i}$ denote total time spent in state $r$. The complete-data log-likelihood is
\begin{equation}
\ell_c(\theta)=
\sum_i\sum_{r\ne s}N_{rs,i}x_i^\top\beta_{rs}
-\sum_i\sum_r T_{r,i}\sum_{s\ne r}\exp(x_i^\top\beta_{rs}).
\label{eq:complete_lik}
\end{equation}
The EM algorithm replaces the unobserved sufficient statistics $(N_{rs,i},T_{r,i})$ by their conditional expectations given the observed panel data.

\subsection{Deterministic E-step}

At iteration $m$, let $Q_i^{(m)}=Q(x_i;\theta^{(m)})$, $R_i^{(m)}=I+Q_i^{(m)}/\lambda_i^{(m)}$, and consider one interval with endpoints $a,b$ and length $t$.

The posterior probability of $K=k$ embedded events is
\begin{equation}
w_k(a,b,t;\theta^{(m)})=
\frac{
 e^{-\lambda_i t}(\lambda_i t)^k/k!\,[(R_i^{(m)})^k]_{ab}
}{
 [\exp\{Q_i^{(m)}t\}]_{ab}
}.
\label{eq:posterior_k}
\end{equation}
Conditional on $K=k$, the embedded Markov bridge probability is
\begin{equation}
P(Z_u=r\mid Z_0=a,Z_k=b,K=k)=
\frac{[(R_i^{(m)})^u]_{ar}[(R_i^{(m)})^{k-u}]_{rb}}
{[(R_i^{(m)})^k]_{ab}}.
\label{eq:bridge_state}
\end{equation}
Therefore, the expected state occupancy time is
\begin{equation}
E(T_r\mid a,b,t)=
\sum_{k=0}^{\infty}w_k(a,b,t;\theta^{(m)})
\sum_{u=0}^{k}\frac{t}{k+1}
P(Z_u=r\mid Z_0=a,Z_k=b,K=k),
\label{eq:expected_time}
\end{equation}
and the expected transition count is
\begin{equation}
E(N_{rs}\mid a,b,t)=
\sum_{k=0}^{\infty}w_k(a,b,t;\theta^{(m)})
\sum_{u=1}^{k}
\frac{[(R_i^{(m)})^{u-1}]_{ar}(R_i^{(m)})_{rs}[(R_i^{(m)})^{k-u}]_{sb}}
{[(R_i^{(m)})^k]_{ab}}.
\label{eq:expected_count}
\end{equation}
In implementation, the infinite sums are truncated once the remaining Poisson tail probability is below a prespecified tolerance.

\subsection{M-step}

After aggregating expected sufficient statistics over all intervals, the transition-specific expected complete-data objective is
\begin{equation}
Q(\beta_{rs})=
\sum_i \widehat N_{rs,i}x_i^\top\beta_{rs}
-
\sum_i \widehat T_{r,i}\exp(x_i^\top\beta_{rs}),
\label{eq:mstep_objective}
\end{equation}
where $\widehat N_{rs,i}=E(N_{rs,i}\mid \mathcal O_i)$ and $\widehat T_{r,i}=E(T_{r,i}\mid \mathcal O_i)$.

Equation~\eqref{eq:mstep_objective} is equivalent to a Poisson regression with mean
\begin{equation}
E(\widehat N_{rs,i})=\widehat T_{r,i}\exp(x_i^\top\beta_{rs}),
\end{equation}
where $\log \widehat T_{r,i}$ acts as an offset. 
Thus each allowed transition can be updated independently using standard Newton--Raphson or 
iteratively reweighted least squares updates.

\subsubsection{Observed-information covariance estimation}
At the final EM estimate, covariance is calculated using Louis' identity rather
than the expected complete-data information alone. Let $S_{c,ij}(\theta)$ and
$H_{c,ij}(\theta)$ denote the complete-path score and Hessian for observation
interval $j$ of subject $i$. The observed information is
\begin{equation}
\widehat{I}_{\mathrm{obs}}
=-\sum_{i,j}E\{H_{c,ij}(\hat\theta)\mid\mathcal O\}
-\sum_{i,j}\left[
E\{S_{c,ij}(\hat\theta)S_{c,ij}(\hat\theta)^\top\mid\mathcal O\}
-\bar S_{ij}\bar S_{ij}^\top\right],
\label{eq:louis_information}
\end{equation}
where $\bar S_{ij}=E\{S_{c,ij}(\hat\theta)\mid\mathcal O\}$.
The first term is the complete information and the second subtracts the
information missing because the path between observed endpoints is latent.
Consequently, Equation~\eqref{eq:louis_information} retains covariance between
coefficients belonging to different transitions.

We use
\begin{equation}
\widehat{\operatorname{Var}}(\hat\theta)
=\widehat I_{\mathrm{obs}}^{-1},\qquad
SE(\hat\theta_l)=
\sqrt{[\widehat I_{\mathrm{obs}}^{-1}]_{ll}},
\end{equation}
and construct Wald intervals as $\hat\theta_l\pm1.96SE(\hat\theta_l)$.
Further computational details are provided in Section~S8 of the
Supplementary Material.

\subsection{Posterior latent trajectory inference}

A key advantage of the proposed deterministic uniformization framework is that it supports posterior inference for clinically meaningful latent pathways without numerical integration over path space or Monte Carlo path sampling. The posterior event-count distribution and embedded Markov bridge are already given in Equations~(\ref{eq:posterior_k}) and~(\ref{eq:bridge_state}); the corresponding expected occupancy times and transition counts in Equations~(\ref{eq:expected_time}) and~(\ref{eq:expected_count}) are sufficient statistics for EM estimation. Exact clinical-pathway inference instead uses the following collapsed-path recursion.

Consider an observation interval

\[
(t_j,t_{j+1}],
\]

with observed endpoints

\[
X(t_j)=a,
\qquad
X(t_{j+1})=b,
\]

and interval length

\[
\Delta t=t_{j+1}-t_j.
\]

Under uniformization, let

\[
K\sim\mathrm{Poisson}(\lambda\Delta t),
\]

together with an embedded discrete-time Markov chain

\[
Z_0,Z_1,\ldots,Z_K,
\]

with

\[
Z_0=a,
\qquad
Z_K=b.
\]

For competing-pathway models, we define a clinical pathway as the sequence of distinct states after deleting uniformization self-transitions. For a candidate pathway $\omega=(s_0,\ldots,s_m)$, let $D_k(j)$ be the probability that, after $k$ embedded-chain steps, the collapsed path equals $(s_0,\ldots,s_j)$ and the current state is $s_j$. Initialize $D_0(0)=1$, and use
\begin{equation}
D_{k+1}(j)=D_k(j)R_{s_js_j}+\mathbf{1}(j>0)D_k(j-1)R_{s_{j-1}s_j},
\label{eq:pathdp}
\end{equation}
with entries outside $0,\ldots,m$ set to zero. The exact posterior probability of the collapsed clinical pathway is then
\begin{equation}
P(\omega\mid a,b,\Delta t)
=
\frac{\sum_{k=0}^{\infty}e^{-\lambda\Delta t}(\lambda\Delta t)^kD_k(m)/k!}
{P_{ab}(\Delta t)}.
\label{eq:exactpathprob}
\end{equation}
The same Poisson truncation used by the E-step makes this a deterministic calculation. It distinguishes clinical state sequences from uniformization sequences, which contain artificial self-transitions. The maximum a posteriori clinical pathway is

\begin{equation}
\hat{\omega}
=
\arg\max_{\omega}
P(\omega\mid Y),
\label{eq:maptrajectory}
\end{equation}

where the maximization is over prespecified scientifically meaningful candidate pathways.

As a computational shortcut, posterior expected branching counts may be normalized, for example
\begin{equation}
\widetilde P(1\to2\to4\mid Y)=
\frac{E(N_{12}\mid Y)}{E(N_{12}\mid Y)+E(N_{13}\mid Y)}.
\label{eq:pathshortcut}
\end{equation}
In general, this expected-count ratio measures the relative posterior use of the branching transitions and need not equal the posterior probability of a complete clinical pathway. 
It is exact only when the candidate pathways are mutually exclusive and exhaustive and each contains its defining transition exactly once. 
For recurrent or overlapping routes, the exact dynamic-programming probability in Equation~(\ref{eq:exactpathprob}) should be used. 

Consequently, the proposed deterministic uniformization framework simultaneously supports efficient parameter estimation and interpretable latent trajectory reconstruction within a unified deterministic inference procedure. Computational details for posterior pathway probability calculations are provided in the Supplementary Material.

\section{Theoretical Properties}
\label{sec:theory}

This section summarizes the main theoretical properties of the proposed deterministic uniformization EM estimator. The theoretical development consists of two parts. The first part focuses on the convergence properties of the EM algorithm, establishing its validity from an algorithmic perspective. The second part investigates the large-sample properties of the resulting estimator, including consistency and asymptotic normality. Together, these results demonstrate the validity of the proposed method from both computational and statistical viewpoints. Full proofs and detailed regularity conditions are provided in the Supplementary Material.

\subsection{Regularity conditions}

Throughout the theoretical development, we assume:

(A1) the parameter space $\Theta$ is compact and the true parameter
$\theta_0$ lies in its interior;

(A2) transition intensities are uniformly bounded away from zero and infinity, observation intervals are uniformly bounded, observed transition probabilities are uniformly positive, and the number of observation intervals per subject is uniformly bounded;

(A3) transition probabilities are continuous in $\theta$;

(A4) the exact normalized log-likelihood satisfies a uniform law of large numbers;

(A5) the observed-data model is identifiable, so that the population objective function has a unique maximizer at $\theta_0$;

(A6) transition probabilities are twice continuously differentiable and the observed-data likelihood satisfies the standard Fisher information regularity conditions.

These assumptions ensure stable likelihood evaluation, finite conditional sufficient statistics, and identifiability of the target parameter. Assumption A2 is primarily used in establishing the convergence properties of the EM algorithm, whereas Assumptions A1 and A3--A6 are employed in deriving the asymptotic properties of the resulting estimator.

\subsection{EM algorithm theory}

\begin{theorem}[Transition probability approximation]
\label{thm:transition}

Under condition A2, transition intensities are uniformly
bounded. Therefore, for each subject \(i\),

\[
\lambda_i
\ge
\sup_{\theta\in\Theta}
\max_r
\{-q_{rr}(x_i;\theta)\}
<
\infty
\]

can be chosen as a finite dominating uniformization rate.

For any states \(a,b\), let

\[
P_{ab}(t;\theta)
=
[e^{Q(\theta)t}]_{ab}
\]

denote the exact transition probability and

\[
P_{K,ab}(t;\theta)
=
\left[
e^{-\lambda_i t}
\sum_{k=0}^{K}
\frac{(\lambda_i t)^k}{k!}
R(\theta)^k
\right]_{ab}
\]

denote its deterministic uniformization approximation.

Then

\[
\sup_{\theta\in\Theta}
\left|
P_{K,ab}(t;\theta)
-
P_{ab}(t;\theta)
\right|
\le
\rho_K(t),
\]

where

\[
\rho_K(t)
=
P(N_i>K),
\qquad
N_i
\sim
\mathrm{Poisson}(\lambda_i t).
\]

Moreover,

\[
\rho_K
\longrightarrow
0
\qquad
(K\to\infty).
\]

\end{theorem}

\begin{theorem}[Uniform approximation of the EM auxiliary function]
\label{thm: Q}
For subsequent sample-level results, define the maximal interval-level transition probability
truncation error

\[
\rho_K
=
\max_{i,j}
\rho_K(\Delta t_{ij}).
\]
Let

\begin{equation}
Q_n(\theta|\theta^{(m)})
=
\sum_{i=1}^n
\sum_{r\neq s}
\widehat N_{rs,i}^{(m)}
\log q_{rs}(x_i;\theta)
-
\sum_{i=1}^n
\sum_r
\widehat T_{r,i}^{(m)}
q_r(x_i;\theta),
\end{equation}

where

\[
\widehat N_{rs,i}^{(m)}
=
E_{\theta^{(m)}}[N_{rs,i}\mid Y_i],
\qquad
\widehat T_{r,i}^{(m)}
=
E_{\theta^{(m)}}[T_{r,i}\mid Y_i].
\]

Let \(Q_{n,K}(\theta|\theta^{(m)})\) denote the same quantity computed with
the \(K\)-term truncated uniformization E-step.

Under condition A2,

\[
\sup_{\theta\in\Theta}
\left|
Q_{n,K}(\theta|\theta^{(m)})
-
Q_n(\theta|\theta^{(m)})
\right|
\le
C_n\rho_{K-1}.
\]

Consequently,

\[
\sup_{\theta\in\Theta}
\left|
Q_{n,K}(\theta|\theta^{(m)})
-
Q_n(\theta|\theta^{(m)})
\right|
\to0
\]

as \(K\to\infty\).
\end{theorem}

\begin{theorem}[Approximate monotonicity under truncation]

Suppose the conditions of Theorem 2 hold. Let

\[
\epsilon_K
=
\sup_{\theta\in\Theta}
\left|
Q_{n,K}(\theta|\theta^{(m)})
-
Q_n(\theta|\theta^{(m)})
\right|.
\]

Then

\begin{equation}
\ell_n(\theta^{(m+1)})
\ge
\ell_n(\theta^{(m)})
-
2\epsilon_K.
\end{equation}

\end{theorem}

Consequently, the deterministic uniformization EM algorithm is an asymptotically monotone generalized EM algorithm.

\subsection{Asymptotic theory}
The preceding approximation results imply that the proposed estimator inherits the same large-sample behavior as the exact maximum likelihood estimator. In particular, consistency and asymptotic normality follow from standard M-estimation arguments together with the uniform approximation established in Theorems 1–3.

\begin{theorem}[Consistency]
\label{thm:consistency}

Suppose conditions
A1-A5
hold.

Let

\[
\hat\theta_{n,K_n}
=
\arg\max_{\theta\in\Theta}
M_{n,K_n}(\theta),
\]

where the truncation level \(K_n\) is chosen so that

\[
\rho_{K_n}\to0.
\]

Then

\[
\hat\theta_{n,K_n}
\overset{p}{\longrightarrow}
\theta_0.
\]

\end{theorem}

\begin{theorem}[Asymptotic normality]
\label{thm:asymptotic_normality}

Suppose the conditions A1-A6 hold.

Assume that
\[
\sqrt n\,\eta_{K_n}
\longrightarrow
0,
\]
where \(\eta_{K_n}\) is the uniform interval-level truncation error.

Then
\[
\sqrt n
\left(
\hat\theta_{n,K_n}
-
\theta_0
\right)
\Rightarrow
N
\left(
0,
I(\theta_0)^{-1}
\right).
\]

\end{theorem}

Detailed proofs follow from standard M-estimation arguments together with the uniform approximation results established above.

In practice, standard errors are computed from the inverse observed-data
information evaluated through Louis' identity. The deterministic
uniformization recursion calculates both the complete-information and
missing-information components without numerical differentiation of the
observed likelihood.

\subsection{Computational complexity}

Direct likelihood evaluation based on repeated matrix exponentials requires
$O(S^3)$ operations per interval, yielding total complexity
$O(nmS^3)$
per likelihood evaluation, where $n$ is the number of subjects, $m$ is the maximum number of intervals, and $S$ is the number of states.
Under deterministic uniformization, recursive probability propagation requires
$O(S^2)$
per embedded step for dense transition structures and
$O(E)$
for sparse structures, where $E$ is the number of allowed transitions.
With truncation level $K$, the E-step complexity is therefore
$O(nmKS^2)$
for dense models and
$O(nmKE)$
for sparse models.
The M-step reduces to transition-specific Poisson regressions and is of lower order when the number of covariates is fixed.

\section{Simulation Study}\label{sec:simulation}

We conducted simulation studies to evaluate estimation accuracy, uncertainty quantification, convergence behavior, 
and computational scalability of the proposed deterministic uniformization EM algorithm. 
The proposed method was compared with direct matrix-exponential likelihood optimization.

Four transition structures were considered.

\paragraph{Scenario A: three-state progressive model.}
The baseline setting used transitions
\[
1\to2,\qquad 1\to3,\qquad 2\to3,
\]
with state 3 absorbing. This setting evaluated performance under relatively simple latent trajectory structure.

\paragraph{Scenario B: four-state progressive model.}
The second setting used
\[
1\to2\to3\to4,
\]
with state 4 absorbing. This scenario introduced longer latent paths while maintaining a simple progressive structure.

\paragraph{Scenario C: competing pathway model.}
The third setting used two competing pathways,
\[
1\to2\to4,
\qquad
1\to3\to4,
\]
with state 4 absorbing. This scenario was designed to evaluate inference under latent trajectory ambiguity while retaining subject-specific covariate information. Let $z_{i2}=x_{i2}-0.5$. The branching intensities were
\begin{equation}
\begin{split}
\log q_{12}(x_i)&=\log(p_0)+0.40x_{i1}+0.05z_{i2},\\
\log q_{13}(x_i)&=\log(1-p_0)-0.25z_{i2},
\end{split}
\end{equation}
so that
\[
\log\{q_{12}(x_i)/q_{13}(x_i)\}
=\operatorname{logit}(p_0)+0.40x_{i1}+0.30z_{i2}.
\]
Low, moderate, and high ambiguity used $p_0=0.95$, $0.70$, and $0.50$, respectively. The covariate contrast was held fixed across ambiguity levels. To prevent post-branch absorption speed from confounding the ambiguity comparison, both downstream transitions used
\[
\log q_{24}(x_i)=\log q_{34}(x_i)=-1.60+0.20x_{i1}-0.10z_{i2}.
\]

\paragraph{Scenario D: one-return competing-pathway network.}
The fourth setting added one return transition to the C-high network:
\[
1\to2,\quad
1\to3,\quad
2\to4,\quad
3\to4,\quad
2\to1.
\]
This setting assessed scalability and numerical stability in the presence of a recurrent pathway structure.
\subsection{Data generation}
For each scenario, we generated 100 independent datasets, each containing 500 subjects.

For all scenarios, transition intensities followed equation~\eqref{eq:intensity} with covariates
\[
x_i=(1,x_{i1},x_{i2})^\top,
\qquad
x_{i1}\sim N(0,1),
\qquad
x_{i2}\sim\mathrm{Bernoulli}(0.5).
\]
For each subject, a full CTMC trajectory was generated exactly from the specified generator matrix. Irregular visit times were then sampled over a fixed follow-up window, and only the states observed at visit times were retained. Intermediate transition times and states were discarded.

\subsection{Methods compared and Performance measures}

We compared the proposed deterministic uniformization EM estimator with maximum likelihood from the 
R package \texttt{msm}, which evaluates interval transition probabilities using matrix exponentials. 
Both methods were run in every scenario. The proposed method used the Louis observed information for standard errors, while \texttt{msm} used its numerical observed-likelihood Hessian. Each method's reported runtime includes point estimation and covariance calculation.

Across repeated datasets, coefficient-level performance was summarized by the true coefficient value, mean estimate, mean estimated standard error, and empirical 95\% confidence interval coverage probability. Method-level performance was summarized by mean coverage across coefficients, mean runtime, and the mean number of EM iterations or \texttt{msm} likelihood evaluations.

For Scenario C, the model was fitted separately to each complete simulated panel dataset. Latent-path reconstruction was then performed for every observation interval whose observed endpoints were states 1 and 4 and whose simulated collapsed path was either $1\to2\to4$ or $1\to3\to4$. The true simulated path was used only as the validation label, not as an input to estimation or posterior calculation. For each eligible interval, Equation~(\ref{eq:exactpathprob}) was evaluated using only its endpoints, length, covariates, and the fitted parameters. The reconstructed pathway \(\hat{\omega}\) was the candidate with the larger posterior probability. We evaluated pathway classification accuracy,

\[
P(\hat{\omega}=\omega),
\]

where \(\omega\) denotes the true latent pathway. We calculated this accuracy separately for the exact dynamic-programming posterior in Equation~(\ref{eq:exactpathprob}) and the expected-count-ratio shortcut in Equation~(\ref{eq:pathshortcut}). For either method, the reconstructed route was $1\to2\to4$ when its probability or score was at least 0.5 and $1\to3\to4$ otherwise. Because the same simulated panel dataset is used for parameter estimation and pathway reconstruction, this analysis evaluates fitted latent-path recovery rather than external predictive performance.

\subsection{Estimation accuracy}

Table~\ref{tab:method_summary} summarizes interval coverage, computation time, and mean iteration counts 
across all six simulation settings. The uniformization EM estimator and \texttt{msm} produced essentially identical empirical coverage in every setting, while the proposed implementation required less computation time.
\begin{table}[ht]
\centering
\caption{Simulation performance over 100 datasets with $N=500$. Iter. is the mean number of EM iterations, 
Eval. is the mean number of \texttt{msm} likelihood evaluations, and 
speedup is the ratio of mean \texttt{msm} time to mean EM time.}
\label{tab:method_summary}
\vspace{1em}
\footnotesize
\setlength{\tabcolsep}{3pt}
\begin{tabular}{lrrrrrrr}
\toprule
Scenario & \multicolumn{3}{c}{Deterministic Uniformization EM} & \multicolumn{3}{c}{R \texttt{msm}} & Speedup \\
\cmidrule(lr){2-4}\cmidrule(lr){5-7}
& Coverage & Time (s) & Iter. & Coverage & Time (s) & Eval. & $t_{\rm msm}/t_{\rm EM}$ \\
\midrule
A & 0.949 & 0.301 & 14.70 & 0.949 & 1.382 & 60.07 & 4.59 \\
B & 0.947 & 0.184 & 5.10 & 0.946 & 3.040 & 106.50 & 16.48 \\
C-low & 0.940 & 0.398 & 15.99 & 0.939 & 14.901 & 324.53 & 37.45 \\
C-moderate & 0.946 & 0.307 & 11.12 & 0.946 & 3.942 & 68.78 & 12.84 \\
C-high & 0.947 & 0.300 & 10.95 & 0.947 & 3.327 & 55.29 & 11.07 \\
D & 0.945 & 0.640 & 17.75 & 0.945 & 7.166 & 112.15 & 11.20 \\
\bottomrule
\end{tabular}
\end{table}

The 95\% Wald confidence intervals based on Louis' observed information and those based on the numerical Hessian from \texttt{msm} had nearly identical empirical coverage.
Under the scenario C settings, average coverage ranged from 0.940 to 0.947 for uniformization EM and from 0.939 
to 0.947 for \texttt{msm}. In one of the 100 Scenario B replicates, \texttt{msm} returned point estimates but 
could not compute standard errors because the numerically evaluated observed-information matrix was not positive definite. 
Consequently, the reported \texttt{msm} coverage for Scenario B was calculated using the 99 replicates with valid standard errors. 
Coefficient-level results are reported in the Supplementary Material.

\begin{table}[!ht]
\centering
\caption{Coefficient estimation results under Scenario C (high ambiguity). Results are averaged over simulation replicates.}
\label{tab:scenarioC_coef}
\footnotesize
\begin{tabular}{llrccc|ccc}
\toprule
&&&
\multicolumn{3}{c}{Deterministic Uniformization EM}
&
\multicolumn{3}{c}{R \texttt{msm}}
\\
\cmidrule(lr){4-6}
\cmidrule(lr){7-9}
Transition &
Coefficient &
True &
Mean est. &
Mean SE &
Coverage &
Mean est. &
Mean SE &
Coverage
\\
\midrule
$1\to2$ & Intercept & -0.718 & -0.698 & 0.103 & 0.920 & -0.698 & 0.103 & 0.920 \\
         & $x_1$     &  0.400 &  0.394 & 0.072 & 0.950 &  0.394 & 0.072 & 0.950 \\
         & $x_2$     &  0.050 &  0.028 & 0.139 & 0.920 &  0.028 & 0.139 & 0.920 \\
\addlinespace

$1\to3$ & Intercept & -0.568 & -0.552 & 0.096 & 0.940 & -0.552 & 0.096 & 0.940 \\
         & $x_1$     &  0.000 &  0.001 & 0.071 & 0.940 &  0.001 & 0.071 & 0.940 \\
         & $x_2$     & -0.250 & -0.284 & 0.139 & 0.980 & -0.283 & 0.139 & 0.980 \\
\addlinespace

$2\to4$ & Intercept & -1.550 & -1.542 & 0.146 & 0.960 & -1.542 & 0.146 & 0.960 \\
         & $x_1$     &  0.200 &  0.181 & 0.102 & 0.960 &  0.181 & 0.102 & 0.960 \\
         & $x_2$     & -0.100 & -0.111 & 0.197 & 0.960 & -0.111 & 0.197 & 0.960 \\
\addlinespace

$3\to4$ & Intercept & -1.550 & -1.549 & 0.135 & 0.950 & -1.549 & 0.135 & 0.950 \\
         & $x_1$     &  0.200 &  0.222 & 0.105 & 0.930 &  0.222 & 0.105 & 0.930 \\
         & $x_2$     & -0.100 & -0.099 & 0.206 & 0.950 & -0.099 & 0.206 & 0.950 \\
\bottomrule
\end{tabular}
\end{table}

Table~\ref{tab:scenarioC_coef} shows close agreement of point estimates, standard errors, and coverage under Scenario C-high. Detailed coefficient-specific results for all scenarios are provided in the Supplementary Material.

Overall, these findings demonstrate that deterministic uniformization EM achieves estimation accuracy comparable to direct likelihood optimization while maintaining reliable uncertainty quantification across a range of latent trajectory structures.

\subsection{Computational efficiency and scalability}

Both methods were run sequentially without parallelism. 
Mean total runtime ranged from 0.184 to 0.640 seconds for uniformization EM and from 1.382 to 14.901 
seconds for \texttt{msm}. The proposed implementation was faster in every scenario, 
with \texttt{msm}/uniformization EM mean-time ratios ranging from 4.59 in Scenario A to 37.45 in the scenario 
C-low setting. C-moderate and C-high yielded speedups of 12.84 and 11.07, respectively. 
All uniformization 
EM fits converged and returned Louis standard errors.

The scenario C-low setting exposed an additional practical advantage of the EM formulation. Because the $1\to3\to4$ pathway has only about 5\% baseline probability, 
information about its transition coefficients is sparse and the observed likelihood 
can be relatively flat in the corresponding directions. 
The median \texttt{msm} runtime was 9.63 seconds, but its 95th percentile reached 62.51 seconds; 
the corresponding uniformization EM values were 0.35 and 0.58 seconds. 
Mean iteration counts were 324.5 likelihood evaluations for \texttt{msm} and 16.0 EM iterations. 
In contrast to direct matrix-exponential optimization, uniformization EM replaces the missing 
rare-path transitions by posterior expected counts and updates each transition through a weighted Poisson regression. This does not remove the statistical uncertainty caused by a rare pathway, but it made the numerical computation substantially more stable and predictable in this experiment. The two methods' maximized log likelihoods agreed to a median absolute difference of $1.1\times10^{-4}$ in C-low; seven of 100 replicates had differences exceeding 0.01, consistent with occasional optimization difficulty in the sparse rare-path setting.

\subsection{Latent trajectory recovery}

A distinguishing feature of the proposed framework is its ability to recover latent trajectory structure in 
addition to estimating transition intensities. Scenario C was specifically designed to evaluate posterior 
latent pathway inference under varying degrees of pathway ambiguity.

Table~\ref{tab:scenarioC_path_prediction} summarizes fitted latent-path reconstruction under 
the scenario C settings. Under low ambiguity, classification accuracy was 94.8\%. Across 100 datasets, 6,109 observed $1\to4$ intervals were eligible.
\begin{table}[ht]
\centering
\caption{Latent-path reconstruction accuracy in Scenario C. Each model was fitted to its complete simulated dataset of 500 subjects; results pool eligible observed $1\to4$ intervals over 100 replicates.}
\label{tab:scenarioC_path_prediction}

\footnotesize
\setlength{\tabcolsep}{3pt}

\begin{tabular}{lccc}
\toprule
Level &
Intervals &
Dynamic programming accuracy &
Count-ratio accuracy \\
\midrule
Low      & 6,109 & 0.948 & 0.948 \\
Moderate & 6,035 & 0.708 & 0.708 \\
High     & 6,313 & 0.556 & 0.556 \\
\bottomrule
\end{tabular}
\end{table}
The ambiguity manipulation produced the intended monotone pattern: 
reconstruction accuracy decreased from 0.948 under low ambiguity to 0.708 under moderate 
ambiguity and 0.556 under high ambiguity. The exact dynamic programming method and expected-count shortcut made identical pathway classifications in every evaluated interval. This agreement is specific to the Scenario C graph, in which the two routes are mutually exclusive and exhaustive and each route contains its branching transition exactly once; it should not be assumed for recurrent or overlapping pathways.

Across all simulation scenarios, the proposed deterministic uniformization EM algorithm and direct likelihood optimization produced closely agreeing mean coefficient estimates, standard errors, and empirical coverage probabilities, while the proposed method required less total computation time. 
Its transition-specific M-step was also less sensitive to sparse rare-path information than direct optimization in Scenario C. In addition to efficient estimation, the framework provided posterior inference for latent progression pathways and posterior summaries of transition counts and state occupancy times. Pathway reconstruction accuracy decreased as the competing routes became less distinguishable. These results support deterministic uniformization as a practical framework for inference and latent trajectory reconstruction in interval-censored multistate models.

\section{Real Data Application}\label{sec:application}

We applied deterministic uniformization EM to the cardiac allograft
vasculopathy (CAV) data distributed with the R package \texttt{msm}
\citep{jackson2011multi}. The data contain 2,846 records from 622 heart-transplant
recipients monitored at approximately annual examinations. States 1, 2, and 3
represent no CAV, mild/moderate CAV, and severe CAV, respectively; state 4 denotes
death. We allowed transitions $1\leftrightarrow2$, $2\leftrightarrow3$, and
$r\to4$ for $r=1,2,3$, with death absorbing.

For each of the 2,224 observation intervals, we specified
\begin{equation}
q_{rs,ij}=\exp\left\{\beta_{rs,0}
+\beta_{rs,1}\frac{\mathrm{age}_{i,j-1}-50}{10}
+\beta_{rs,2}\mathrm{cumrej}_{i,j-1}\right\},
\label{eq:cav_intensity}
\end{equation}
where $\mathrm{cumrej}$ is the cumulative number of rejection episodes.
The two covariates were held constant at their values measured at the beginning
of each interval. This extends the subject-constant specification to
piecewise-constant intensities by using the corresponding $Q_{ij}$ separately
in each interval's E-step. The model has 21 parameters: an intercept and two
covariate effects for each of seven transitions. Both implementations used the
same transition graph and covariates, and conditioned on the initial observed
state. 

Both fits converged. The uniformization EM and \texttt{msm} log likelihoods were
$-1958.308560$ and $-1958.308568$, respectively. EM required 66 iterations, while
\texttt{msm} required 110 optimizer function evaluations.
The complete uniformization EM analysis, including Louis observed-information standard errors,
required 1.331 seconds, compared with 6.657 seconds for \texttt{msm}.The total cost time ratio was
approximately 5.00 in favor of the proposed method.

Table~\ref{tab:cav_application} presents all estimated log-intensity
coefficients. For example, the EM age coefficient for $1\to4$ was 0.624,
corresponding to an intensity ratio of approximately $\exp(0.624)=1.87$ per
additional decade, conditional on rejection count. The rejection-count
coefficient for $1\to2$ was 0.127, corresponding to an intensity ratio of
approximately 1.14 per additional episode.

\begin{table}[p]
\centering
\small
\setlength{\tabcolsep}{4pt}
\caption{CAV log-intensity coefficients and 95\% Wald intervals. Age is
centered at 50 years and scaled per decade; rejection count is measured per
episode. Both methods use observed-information Wald intervals.}
\label{tab:cav_application}
\begin{tabular}{llcc}
\toprule
Transition & Coefficient & Deterministic Uniformization EM & R \texttt{msm} \\
\midrule
1 $\to$ 2 & Intercept & -2.190 (-2.386, -1.995) & -2.190 (-2.386, -1.995) \\
1 $\to$ 2 & $(\mathrm{Age}-50)/10$ & 0.161 (0.028, 0.293) & 0.161 (0.028, 0.293) \\
1 $\to$ 2 & Rejections & 0.127 (0.059, 0.194) & 0.127 (0.060, 0.194) \\
1 $\to$ 4 & Intercept & -3.001 (-3.240, -2.761) & -3.001 (-3.240, -2.761) \\
1 $\to$ 4 & $(\mathrm{Age}-50)/10$ & 0.624 (0.341, 0.908) & 0.624 (0.341, 0.907) \\
1 $\to$ 4 & Rejections & -0.054 (-0.199, 0.091) & -0.054 (-0.199, 0.091) \\
2 $\to$ 1 & Intercept & -1.211 (-1.677, -0.745) & -1.211 (-1.677, -0.745) \\
2 $\to$ 1 & $(\mathrm{Age}-50)/10$ & 0.224 (-0.108, 0.555) & 0.224 (-0.108, 0.556) \\
2 $\to$ 1 & Rejections & -0.087 (-0.271, 0.096) & -0.087 (-0.271, 0.096) \\
2 $\to$ 3 & Intercept & -1.201 (-1.534, -0.867) & -1.201 (-1.535, -0.867) \\
2 $\to$ 3 & $(\mathrm{Age}-50)/10$ & -0.263 (-0.517, -0.010) & -0.263 (-0.517, -0.010) \\
2 $\to$ 3 & Rejections & -0.003 (-0.117, 0.112) & -0.002 (-0.117, 0.112) \\
2 $\to$ 4 & Intercept & -2.985 (-3.801, -2.170) & -2.985 (-3.801, -2.170) \\
2 $\to$ 4 & $(\mathrm{Age}-50)/10$ & 0.052 (-0.521, 0.625) & 0.052 (-0.520, 0.624) \\
2 $\to$ 4 & Rejections & 0.205 (0.014, 0.397) & 0.206 (0.014, 0.397) \\
3 $\to$ 2 & Intercept & -1.471 (-2.199, -0.743) & -1.472 (-2.200, -0.744) \\
3 $\to$ 2 & $(\mathrm{Age}-50)/10$ & -0.609 (-1.158, -0.061) & -0.609 (-1.157, -0.061) \\
3 $\to$ 2 & Rejections & -0.192 (-0.491, 0.107) & -0.192 (-0.491, 0.108) \\
3 $\to$ 4 & Intercept & -1.344 (-1.762, -0.926) & -1.343 (-1.761, -0.926) \\
3 $\to$ 4 & $(\mathrm{Age}-50)/10$ & -0.118 (-0.413, 0.177) & -0.118 (-0.413, 0.177) \\
3 $\to$ 4 & Rejections & 0.084 (-0.025, 0.193) & 0.084 (-0.025, 0.193) \\
\bottomrule
\end{tabular}
\end{table}

We additionally examined the 148 consecutive observation intervals with
endpoints in states 1 and 4. For each interval, the exact dynamic-programming
calculation gave posterior probabilities for the selected clinical pathways
$1\to4$, $1\to2\to4$, and $1\to2\to3\to4$. Because the fitted model permits
the return transitions $2\to1$ and $3\to2$, the remaining posterior
probability was retained as an other/recurrent category. The selected three
paths accounted for 97.4\% of posterior probability on average. Table~\ref{tab:cav_pathway_summary}
classifies each interval by the category with the largest posterior
probability. The direct $1\to4$ path was most probable for 141 intervals
(95.3\%), $1\to2\to4$ for 4 intervals (2.7\%), and
$1\to2\to3\to4$ for 3 intervals (2.0\%). The patient-level probabilities are
reported in the Supplementary Material.

\begin{table}[ht]
\centering
\caption{Posterior classification of the 148 observed CAV intervals with
endpoints in states 1 and 4. Each interval is assigned to the category with
the largest posterior probability.}
\label{tab:cav_pathway_summary}
\footnotesize
\vspace{0.5em}
\begin{tabular}{lrr}
\toprule
Posterior pathway category & Intervals, $n$ & Percent \\
\midrule
$1\to4$ & 141 & 95.3\% \\
$1\to2\to4$ & 4 & 2.7\% \\
$1\to2\to3\to4$ & 3 & 2.0\% \\
Other/recurrent & 0 & 0.0\% \\
\midrule
Total & 148 & 100.0\% \\
\bottomrule
\end{tabular}
\end{table}

\section{Discussion}\label{sec:discussion}

We developed a deterministic uniformization EM algorithm for interval-censored multistate transition models. 
The proposed approach avoids repeated matrix exponential evaluations, 
reduces the M-step to a series of transition-specific Poisson regressions, 
and naturally provides posterior summaries of latent trajectories. 
Theoretical results establish uniform truncation accuracy and approximate EM monotonicity, 
while standard large-sample properties, including consistency and asymptotic normality, 
follow under regularity and identifiability conditions.

The transition-specific EM structure is particularly useful when one pathway is rare. 
Direct observed-likelihood optimization must search over weakly informed coefficient directions and 
may require many matrix-exponential evaluations, 
whereas the EM algorithm updates those coefficients from posterior expected transition counts. 
The resulting computation can be more stable even though the inferential uncertainty associated with 
the rare transition remains and should be reported.

The primary motivation for this framework is its potential application in personalized medicine. 
Patients with the same observed disease stage may follow different underlying pathological pathways 
and exhibit heterogeneous disease progression over time. 
By recovering posterior latent trajectories and underlying disease states, 
the proposed framework provides a statistical foundation for individualized risk assessment and 
may support future development of dynamic treatment strategies and precision medicine applications. 
This capability is particularly relevant when patients with similar observed outcomes may 
follow distinct latent progression pathways.

The CAV application illustrates implementation with covariates held constant at their interval-start 
values based on piecewise constant harzard model. 
The proposed estimator and \texttt{msm} produced closely agreeing point estimates and standard errors, 
with shorter total computation time for the compiled EM implementation in this run. 
Future work includes covariate processes evolving within observation intervals and more complex transition structures.

\section*{Supplementary Material}

The Supplementary Material is organized as follows. Section S1 provides additional details of the deterministic uniformization EM algorithm. Sections S2--S5 establish the main theoretical properties of the proposed method, including regularity conditions, transition probability approximation, uniform approximation of the EM auxiliary function, and approximate monotonicity. Section S6 presents the asymptotic theory, including consistency and asymptotic normality. Section S7 reports additional simulation results, Section S8 provides implementation details, and Section S9 reports patient-level CAV pathway probabilities.

\bibliographystyle{apalike}
\bibliography{references}

\end{document}


\maketitle

\tableofcontents

\newpage

\section{Posterior Latent Trajectory Inference}
\label{sec:supp_trajectory}

This section provides additional details on posterior latent trajectory inference under the deterministic uniformization framework. In addition to parameter estimation, the proposed method yields exact posterior expectations of latent transition counts, state occupancy times, and pathway probabilities without requiring explicit path enumeration or Monte Carlo path sampling.

\subsection{Posterior distribution of the latent event count}

Consider an observation interval of length

\[
\Delta t=t_{j+1}-t_j,
\]

with observed endpoints

\[
X(t_j)=a,
\qquad
X(t_{j+1})=b.
\]

Under uniformization,

\[
K\sim\mathrm{Poisson}(\lambda\Delta t),
\]

where \(K\) denotes the number of latent Poisson events.

By Bayes' theorem,

\begin{equation}
P(K=k\mid a,b,\Delta t)
=
\frac{
P(a,b,K=k)
}{
P(a,b)
}.
\end{equation}

Using the uniformization representation,

\[
P(a,b,K=k)
=
e^{-\lambda\Delta t}
\frac{(\lambda\Delta t)^k}{k!}
[(R)^k]_{ab},
\]

and therefore

\begin{equation}
P(K=k\mid a,b,\Delta t)
=
\frac{
e^{-\lambda\Delta t}
(\lambda\Delta t)^k
[(R)^k]_{ab}/k!
}
{
P_{ab}(\Delta t)
},
\label{eq:supp_postK}
\end{equation}

where

\[
P_{ab}(\Delta t)
=
[\exp(Q\Delta t)]_{ab}.
\]

\subsection{Embedded Markov bridge distribution}

Conditional on

\[
Z_0=a,
\qquad
Z_K=b,
\]

the Markov property implies

\begin{align}
&P(Z_u=r\mid Z_0=a,Z_K=b,K=k)
\\
&=
\frac{
P(Z_u=r,Z_K=b\mid Z_0=a,K=k)
}{
P(Z_K=b\mid Z_0=a,K=k)
}
\\
&=
\frac{
[(R)^u]_{ar}
[(R)^{k-u}]_{rb}
}
{
[(R)^k]_{ab}
},
\end{align}

for

\[
u=0,\ldots,k.
\]

Hence,

\begin{equation}
P(Z_u=r\mid Z_0=a,Z_K=b,K=k)
=
\frac{
[(R)^u]_{ar}
[(R)^{k-u}]_{rb}
}
{
[(R)^k]_{ab}
}.
\label{eq:supp_bridge}
\end{equation}

\subsection{Posterior expected transition counts}

For an observation interval with endpoints

\[
X(0)=a,
\qquad
X(t)=b,
\]

define

\[
N_{rs}
=
\text{number of transitions from state }r\text{ to state }s
\]

occurring during the interval.

Conditional on the latent Poisson count \(K=k\), the expected number of
\(r\rightarrow s\) transitions can be written as

\begin{equation}
E(N_{rs}\mid a,b,K=k,t)
=
\sum_{u=1}^{k}
P(Z_{u-1}=r,Z_u=s\mid a,b,K=k).
\end{equation}

By the embedded Markov bridge representation,

\begin{equation}
P(Z_{u-1}=r,Z_u=s\mid a,b,K=k)
=
\frac{
[(R)^{u-1}]_{ar}
R_{rs}
[(R)^{k-u}]_{sb}
}{
[(R)^k]_{ab}
}.
\end{equation}

Therefore,

\begin{equation}
E(N_{rs}\mid a,b,K=k,t)
=
\sum_{u=1}^{k}
\frac{
[(R)^{u-1}]_{ar}
R_{rs}
[(R)^{k-u}]_{sb}
}{
[(R)^k]_{ab}
}.
\end{equation}

Averaging over the posterior distribution of the latent event count yields

\begin{equation}
E(N_{rs}\mid a,b,t)
=
\sum_{k=0}^{\infty}
P(K=k\mid a,b,t)
\sum_{u=1}^{k}
\frac{
[(R)^{u-1}]_{ar}
R_{rs}
[(R)^{k-u}]_{sb}
}{
[(R)^k]_{ab}
}.
\label{eq:supp_EN}
\end{equation}

\subsection{Posterior expected state occupancy times}

For an observation interval of length \(t\), define

\[
T_r
=
\int_0^t
I\{X(s)=r\}\,ds,
\]

which represents the total time spent in state \(r\).

Conditional on \(K=k\), the \(k+1\) waiting times between successive Poisson events are exchangeable and jointly follow a Dirichlet distribution after conditioning on the total interval length \(t\). Consequently, each segment has expected duration

\[
E(W_u\mid K=k,t)
=
\frac{t}{k+1},
\qquad
u=0,\ldots,k.
\]

Therefore,

\begin{equation}
E(T_r\mid a,b,K=k,t)
=
\sum_{u=0}^{k}
\frac{t}{k+1}
P(Z_u=r\mid a,b,K=k),
\end{equation}

where

\[
P(Z_u=r\mid a,b,K=k)
=
\frac{
[(R)^u]_{ar}
[(R)^{k-u}]_{rb}
}{
[(R)^k]_{ab}
}.
\]

Averaging over the posterior distribution of the latent event count yields

\begin{equation}
E(T_r\mid a,b,t)
=
\sum_{k=0}^{\infty}
P(K=k\mid a,b,t)
\sum_{u=0}^{k}
\frac{t}{k+1}
P(Z_u=r\mid a,b,K=k).
\label{eq:supp_ET}
\end{equation}

\subsection{Posterior pathway probabilities}

Let a clinical pathway $\omega=(s_0,\ldots,s_m)$ be the sequence remaining after all self-transitions are removed from a uniformized embedded-chain realization. Define $D_k(j)$ as the joint probability that after $k$ embedded steps the collapsed sequence is $(s_0,\ldots,s_j)$ and its current state is $s_j$. Starting from $D_0(0)=1$, $D_0(j)=0$ for $j>0$, compute
\begin{equation}
D_{k+1}(0)=D_k(0)R_{s_0s_0},
\end{equation}
and, for $j=1,\ldots,m$,
\begin{equation}
D_{k+1}(j)=D_k(j)R_{s_js_j}+D_k(j-1)R_{s_{j-1}s_j}.
\label{eq:supp_path_dp}
\end{equation}
Transitions outside the candidate sequence contribute zero. Thus
\begin{equation}
P(\omega,a\to b;t)=\sum_{k=0}^{\infty}e^{-\lambda t}
\frac{(\lambda t)^k}{k!}D_k(m),
\end{equation}
and the exact collapsed-path posterior is
\begin{equation}
P(\omega\mid X(0)=a,X(t)=b)=
\frac{P(\omega,a\to b;t)}{P_{ab}(t)}.
\label{eq:supp_exact_path}
\end{equation}
The denominator is evaluated by the same truncated uniformization sum, ensuring that numerator and denominator use an identical numerical tolerance. A single forward recursion over $k=0,\ldots,K$ costs $O(Km)$ for one candidate and does not enumerate uniformization sequences.

The maximum a posteriori pathway is defined as

\begin{equation}
\hat\omega
=
\arg\max_{\omega}
P(\omega\mid Y),
\end{equation}

where \(\omega\) denotes a candidate latent trajectory.

For quick screening, one may instead normalize posterior expected branching counts. For Scenario C this gives

\begin{equation}
\widetilde P(1\rightarrow2\rightarrow4\mid Y)
=
\frac{
E(N_{12}\mid Y)
}{
E(N_{12}\mid Y)+E(N_{13}\mid Y)
},
\end{equation}

and

\begin{equation}
\widetilde P(1\rightarrow3\rightarrow4\mid Y)
=
\frac{
E(N_{13}\mid Y)
}{
E(N_{12}\mid Y)+E(N_{13}\mid Y)
}.
\end{equation}

These ratios equal the exact probabilities in Equation~(\ref{eq:supp_exact_path}) for Scenario C: the two routes are mutually exclusive and exhaustive conditional on an observed $1\to4$ interval, each route uses its branching transition once, and the graph has no return transitions. This equivalence does not generally hold with recurrent transitions, repeated branch crossings, overlapping candidate routes, or omitted routes; in those settings the ratios are descriptive scores and the dynamic program should be used.

For more complex networks with many possible routes, the same recursion can be extended to a finite-state automaton representing a collection of pathway patterns. If explicit pathway probabilities are not scientifically interpretable, the posterior expectations in Equations~(\ref{eq:supp_EN})--(\ref{eq:supp_ET}) provide lower-dimensional summaries through transition counts and occupancy times.

Consequently, deterministic uniformization supports interpretable latent trajectory inference without explicit path enumeration, numerical integration over the latent path space, or Monte Carlo path sampling.

\section{Regularity Conditions}

\begin{assumption}[Compactness]
\label{ass:compact}

The parameter space \(\Theta\) is compact and the true parameter
\(\theta_0\) lies in the interior of \(\Theta\).

\end{assumption}

\begin{assumption}[Model regularity]
\label{ass:model}

Transition intensities are uniformly bounded away from zero and infinity.

Observation intervals are uniformly bounded.

Observed transition probabilities are uniformly positive.

The number of observation intervals per subject is uniformly bounded.

\end{assumption}

\begin{assumption}[Continuity]
\label{ass:continuity}

Transition probabilities are continuous in \(\theta\).

\end{assumption}

\begin{assumption}[Uniform law of large numbers]
\label{ass:ulln}

\[
\sup_{\theta\in\Theta}
|M_n(\theta)-M(\theta)|
\overset{p}{\to}
0.
\]

\end{assumption}

\begin{assumption}[Identifiability]
\label{ass:identifiability}

The population objective function \(M(\theta)\) has a unique maximizer
at \(\theta_0\).

\end{assumption}

\begin{assumption}[Likelihood regularity]
\label{ass:regularity}

Transition probabilities are twice continuously differentiable and the
observed-data likelihood satisfies standard Fisher information
regularity conditions.

\end{assumption}

\section{Transition Probability Approximation}

\begin{theorem}[Transition probability approximation]
\label{thm:transition}

Under Assumption~\ref{ass:model}, transition intensities are uniformly
bounded. Therefore, for each subject \(i\),

\[
\lambda_i
\ge
\sup_{\theta\in\Theta}
\max_r
\{-q_{rr}(x_i;\theta)\}
<
\infty
\]

can be chosen as a finite dominating uniformization rate.

For any states \(a,b\), let

\[
P_{ab}(t;\theta)
=
[e^{Q(\theta)t}]_{ab}
\]

denote the exact transition probability and

\[
P_{K,ab}(t;\theta)
=
\left[
e^{-\lambda_i t}
\sum_{k=0}^{K}
\frac{(\lambda_i t)^k}{k!}
R(\theta)^k
\right]_{ab}
\]

denote its deterministic uniformization approximation.

Then

\begin{equation}
\sup_{\theta\in\Theta}
\left|
P_{K,ab}(t;\theta)
-
P_{ab}(t;\theta)
\right|
\le
\rho_K(t),
\end{equation}

where

\[
\rho_K(t)
=
P(N_i>K),
\qquad
N_i
\sim
\mathrm{Poisson}(\lambda_i t).
\]

Moreover,

\[
\rho_K
\longrightarrow
0
\qquad
(K\to\infty).
\]

\end{theorem}

\begin{proof}

The uniformization representation gives

\[
P_{ab}(t;\theta)
=
\left[
e^{-\lambda_i t}
\sum_{k=0}^{\infty}
\frac{(\lambda_i t)^k}{k!}
R(\theta)^k
\right]_{ab}.
\]

Hence

\[
P_{ab}(t;\theta)
-
P_{K,ab}(t;\theta)
=
\left[
e^{-\lambda_i t}
\sum_{k=K+1}^{\infty}
\frac{(\lambda_i t)^k}{k!}
R(\theta)^k
\right]_{ab}.
\]

Since \(R(\theta)\) is a stochastic matrix,

\[
0
\le
[R(\theta)^k]_{ab}
\le
1.
\]

Therefore,

\[
\left|
P_{ab}(t;\theta)
-
P_{K,ab}(t;\theta)
\right|
\le
e^{-\lambda_i t}
\sum_{k=K+1}^{\infty}
\frac{(\lambda_i t)^k}{k!}.
\]

The right-hand side is exactly the upper tail probability of a
Poisson random variable with mean \(\lambda_i t\):

\[
e^{-\lambda_i t}
\sum_{k=K+1}^{\infty}
\frac{(\lambda_i t)^k}{k!}
=
P(N_i>K).
\]

Consequently,

\[
\sup_{\theta\in\Theta}
\left|
P_{K,ab}(t;\theta)
-
P_{ab}(t;\theta)
\right|
\le
\rho_K(t).
\]

Since \(N_i\) has finite mean \(\lambda_i t\),
its upper-tail probability satisfies

\[
\rho_K(t)
=
P(N_i>K)
\longrightarrow
0
\qquad
(K\to\infty).
\]

This completes the proof.

\end{proof}

\section{Uniform Approximation of the EM Auxiliary Function}
We now show that the uniformization truncation error induces a vanishing
error in the EM auxiliary function. This result justifies the use of the
truncated uniformization E-step in the proposed EM algorithm.

For a single observation interval with endpoints \(a\) and \(b\) and length
\(t\), define the exact bridge probability

\[
w_k(a,b,t;\theta)
=
P_\theta(K=k\mid X(0)=a,X(t)=b),
\]

where \(K\) denotes the number of embedded uniformization events. Under
uniformization,

\[
w_k(a,b,t;\theta)
=
\frac{
e^{-\lambda t}
(\lambda t)^k/k!
[R(\theta)^k]_{ab}
}{
P_{ab}(t;\theta)
}.
\]

The truncated version is

\[
w_{k,K}(a,b,t;\theta)
=
\frac{
e^{-\lambda t}
(\lambda t)^k/k!
[R(\theta)^k]_{ab}
}{
P_{K,ab}(t;\theta)
},
\qquad
k=0,\ldots,K.
\]

\begin{lemma}[Approximation of bridge probabilities]
   Under Assumption~\ref{ass:model},

\begin{equation}
\sum_{k=0}^{K}
\left|
w_{k,K}(a,b,t;\theta)
-
w_k(a,b,t;\theta)
\right|
+
\sum_{k=K+1}^{\infty}
w_k(a,b,t;\theta)
\le
C\rho_K(t),
\end{equation}

uniformly in \(\theta\in\Theta\), for some constant \(C<\infty\). 
\end{lemma}

\begin{proof}

 Let
\[
A_k(\theta)
=
e^{-\lambda t}
\frac{(\lambda t)^k}{k!}
[R(\theta)^k]_{ab}.
\]

Then
\[
P_{ab}(t;\theta)
=
\sum_{k=0}^{\infty}A_k(\theta),
\qquad
P_{K,ab}(t;\theta)
=
\sum_{k=0}^{K}A_k(\theta).
\]

By Theorem~\ref{thm:transition},
\[
0\le
P_{ab}(t;\theta)-P_{K,ab}(t;\theta)
\le
\rho_K(t).
\]

By Assumption~\ref{ass:model},
\[
P_{ab}(t;\theta)\ge c_P>0.
\]

Hence, for sufficiently large \(K\),
\[
P_{K,ab}(t;\theta)
\ge
P_{ab}(t;\theta)-\rho_K(t)
\ge
c_P-\rho_K(t)
\ge
\frac{c_P}{2}.
\]

For \(k\le K\),
\begin{align*}
\left|
w_{k,K}-w_k
\right|
&=
A_k(\theta)
\left|
\frac{1}{P_{K,ab}(t;\theta)}
-
\frac{1}{P_{ab}(t;\theta)}
\right|
\\
&=
A_k(\theta)
\frac{
P_{ab}(t;\theta)-P_{K,ab}(t;\theta)
}{
P_{K,ab}(t;\theta)P_{ab}(t;\theta)
}
\\
&\le
C A_k(\theta)\rho_K(t).
\end{align*}

Summing over \(k=0,\ldots,K\),
\begin{align*}
\sum_{k=0}^{K}
|w_{k,K}-w_k|
&\le
C\rho_K(t)
\sum_{k=0}^{K}A_k(\theta)
\\
&=
C\rho_K(t)P_{K,ab}(t;\theta)
\\
&\le
C\rho_K(t).
\end{align*}

Moreover,
\begin{align*}
\sum_{k=K+1}^{\infty}w_k
&=
\frac{
\sum_{k=K+1}^{\infty}A_k(\theta)
}{
P_{ab}(t;\theta)
}
\\
&\le
\frac{\rho_K(t)}{c_P}
\\
&\le
C\rho_K(t).
\end{align*}

Combining the two bounds gives
\[
\sum_{k=0}^{K}
|w_{k,K}-w_k|
+
\sum_{k=K+1}^{\infty}w_k
\le
C\rho_K(t).
\]

This completes the proof.
\(\square\)
   
\end{proof}

\begin{lemma}[Approximation of expected sufficient statistics]
    Let \(N_{rs}\) denote the number of true CTMC transitions from state \(r\)
to state \(s\) during the interval, and let \(T_r\) denote the total time
spent in state \(r\). Under Assumption~\ref{ass:model},

\begin{equation}
    \left|
E_{\theta,K}
[
N_{rs}
\mid
a,b,t
]
-
E_{\theta}
[
N_{rs}
\mid
a,b,t
]
\right|
\le C\rho_{K-1}(t),
\end{equation}

and

\begin{equation}
    \left|
E_{\theta,K}
[
T_r
\mid
a,b,t
]
-
E_{\theta}
[
T_r
\mid
a,b,t
]
\right|
\le
C\rho_K(t),
\end{equation}

uniformly in \(\theta\in\Theta\).
\end{lemma}

\begin{proof}
 For the occupancy time, write
\[
E_\theta[T_r\mid a,b,t]
=
\sum_{k=0}^{\infty}
E_\theta[T_r\mid a,b,t,K=k]w_k
\]
and
\[
E_{\theta,K}[T_r\mid a,b,t]
=
\sum_{k=0}^{K}
E_\theta[T_r\mid a,b,t,K=k]w_{k,K}.
\]

Since \(0\le T_r\le t\le T_{\max}\), we have
\[
0\le
E_\theta[T_r\mid a,b,t,K=k]
\le T_{\max}.
\]
Therefore,
\begin{align*}
&
\left|
E_{\theta,K}
[
T_r
\mid
a,b,t
]
-
E_{\theta}
[
T_r
\mid
a,b,t
]
\right|
\\
&\le
T_{\max}
\left\{
\sum_{k=0}^{K}
\left|
w_{k,K}-w_k
\right|
+
\sum_{k=K+1}^{\infty}w_k
\right\}.
\end{align*}

By Lemma 1,
\[
\sum_{k=0}^{K}
\left|
w_{k,K}-w_k
\right|
+
\sum_{k=K+1}^{\infty}w_k
\le
C\rho_K(t).
\]
Hence
\[
\left|
E_{\theta,K}
[
T_r
\mid
a,b,t
]
-
E_{\theta}
[
T_r
\mid
a,b,t
]
\right|
\le
C\rho_K(t).
\]

For the transition count, the exact expectation satisfies
\[
E_\theta[N_{rs}\mid a,b,t]
=
\sum_{k=0}^{\infty}
E_\theta[N_{rs}\mid a,b,t,K=k]w_k.
\]
The truncated approximation is
\[
E_{\theta,K}[N_{rs}\mid a,b,t]
=
\sum_{k=0}^{K}
E_\theta[N_{rs}\mid a,b,t,K=k]w_{k,K}.
\]

Taking differences gives
\begin{align*}
&
\left|
E_{\theta,K}[N_{rs}\mid a,b,t]
-
E_{\theta}[N_{rs}\mid a,b,t]
\right|
\\
&\le
\sum_{k=0}^{K}
E_\theta[N_{rs}\mid a,b,t,K=k]
|w_{k,K}-w_k|
\\
&\quad
+
\sum_{k=K+1}^{\infty}
E_\theta[N_{rs}\mid a,b,t,K=k]
w_k.
\end{align*}

Conditional on \(K=k\), the number of true CTMC transitions from
state \(r\) to state \(s\) is at most the number of embedded events.
Thus,
\[
0\le
E_\theta[N_{rs}\mid a,b,t,K=k]
\le k.
\]
Therefore,
\begin{align*}
&
\left|
E_{\theta,K}[N_{rs}\mid a,b,t]
-
E_\theta[N_{rs}\mid a,b,t]
\right|
\\
&\le
\sum_{k=0}^{K}
k|w_{k,K}-w_k|
+
\sum_{k=K+1}^{\infty}
k w_k .
\end{align*}

For the tail term,
\begin{align*}
\sum_{k=K+1}^{\infty} k w_k
&=
\sum_{k=K+1}^{\infty}
k
\frac{
e^{-\lambda t}(\lambda t)^k [R(\theta)^k]_{ab}/k!
}{
P_{ab}(t;\theta)
}
\\
&\le
\frac{1}{c_P}
e^{-\lambda t}
\sum_{k=K+1}^{\infty}
k\frac{(\lambda t)^k}{k!}
\\
&=
\frac{\lambda t}{c_P}
e^{-\lambda t}
\sum_{k=K}^{\infty}
\frac{(\lambda t)^k}{k!}
\\
&\le
C\rho_{K-1}(t),
\end{align*}
where the last inequality uses \(t\le T_{\max}\) and the uniform boundedness
of \(\lambda\).

For the retained terms, using the same bound as in Lemma 1,
\[
|w_{k,K}-w_k|
\le
C A_k(\theta)\rho_K(t).
\]
Thus,
\begin{align*}
\sum_{k=0}^{K} k|w_{k,K}-w_k|
&\le
C\rho_K(t)
\sum_{k=0}^{K} k A_k(\theta)
\\
&\le
C\rho_K(t)
e^{-\lambda t}
\sum_{k=0}^{K}
k\frac{(\lambda t)^k}{k!}
\\
&\le
C\rho_K(t)\lambda t
\\
&\le
C\rho_K(t)
\\
&\le
C\rho_{K-1}(t),
\end{align*}
because \(\rho_K(t)\le\rho_{K-1}(t)\).

Combining the retained-term and tail-term bounds gives
\[
\left|
E_{\theta,K}[N_{rs}\mid a,b,t]
-
E_\theta[N_{rs}\mid a,b,t]
\right|
\le
C\rho_{K-1}(t).
\]

This completes the proof.
\(\square\)

Thus the expected residence times are approximated with error
\(O(\rho_K(t))\), while the expected transition counts are approximated
with error \(O(\rho_{K-1}(t))\).   
\end{proof}

\begin{theorem}[Uniform approximation of the EM auxiliary function]
\label{thm: Q}
For subsequent sample-level results, define the maximal interval-level transition probability
truncation error

\[
\rho_K
=
\max_{i,j}
\rho_K(\Delta t_{ij}).
\]
Let

\[
Q_n(\theta|\theta^{(m)})
=
\sum_{i=1}^n
\sum_{r\neq s}
\widehat N_{rs,i}^{(m)}
\log q_{rs}(x_i;\theta)
-
\sum_{i=1}^n
\sum_r
\widehat T_{r,i}^{(m)}
q_r(x_i;\theta),
\]

where

\[
\widehat N_{rs,i}^{(m)}
=
E_{\theta^{(m)}}[N_{rs,i}\mid Y_i],
\qquad
\widehat T_{r,i}^{(m)}
=
E_{\theta^{(m)}}[T_{r,i}\mid Y_i].
\]

Let \(Q_{n,K}(\theta|\theta^{(m)})\) denote the same quantity computed with
the \(K\)-term truncated uniformization E-step.

Under Assumption~\ref{ass:model},

\[
\sup_{\theta\in\Theta}
\left|
Q_{n,K}(\theta|\theta^{(m)})
-
Q_n(\theta|\theta^{(m)})
\right|
\le
C_n\rho_{K-1}.
\]

Consequently,

\[
\sup_{\theta\in\Theta}
\left|
Q_{n,K}(\theta|\theta^{(m)})
-
Q_n(\theta|\theta^{(m)})
\right|
\to0
\]

as \(K\to\infty\).
\end{theorem}

\begin{proof}
    By the boundedness assumption,

\[
|\log q_{rs}(x_i;\theta)|
\le
C,
\]

and

\[
q_r(x_i;\theta)
=
\sum_{s\neq r}q_{rs}(x_i;\theta)
\le
C.
\]

Therefore,

\begin{align*}
&
\left|
Q_{n,K}(\theta|\theta^{(m)})
-
Q_n(\theta|\theta^{(m)})
\right|
\\
&\le
\sum_{i=1}^n
\sum_{r\neq s}
\left|
\widehat N_{rs,i,K}^{(m)}
-
\widehat N_{rs,i}^{(m)}
\right|
\cdot
|\log q_{rs}(x_i;\theta)|
\\
&\quad
+
\sum_{i=1}^n
\sum_r
\left|
\widehat T_{r,i,K}^{(m)}
-
\widehat T_{r,i}^{(m)}
\right|
q_r(x_i;\theta)
\\
&\le
C
\sum_{i=1}^n
\sum_{r\neq s}
\left|
\widehat N_{rs,i,K}^{(m)}
-
\widehat N_{rs,i}^{(m)}
\right|
+
C
\sum_{i=1}^n
\sum_r
\left|
\widehat T_{r,i,K}^{(m)}
-
\widehat T_{r,i}^{(m)}
\right|.
\end{align*}

Each subject-level statistic is the sum of interval-level statistics.
Applying Lemma 2 to every observed interval gives

\[
\left|
\widehat N_{rs,i,K}^{(m)}
-
\widehat N_{rs,i}^{(m)}
\right|
\le
C m_i\rho_{K-1},
\]

and

\[
\left|
\widehat T_{r,i,K}^{(m)}
-
\widehat T_{r,i}^{(m)}
\right|
\le
C m_i\rho_K.
\]

Since \(m_i\le m_{\max}<\infty\),

\begin{align*}
&
\left|
Q_{n,K}(\theta|\theta^{(m)})
-
Q_n(\theta|\theta^{(m)})
\right|
\\
&\le
C
\sum_{i=1}^n
m_i
\left(
\rho_{K-1}
+
\rho_K
\right)
\\
&\le
C_n
\left(
\rho_{K-1}
+
\rho_K
\right)
\\
&\le
C_n\rho_{K-1},
\end{align*}

The bound is uniform in \(\theta\). Therefore,

\[
\sup_{\theta\in\Theta}
\left|
Q_{n,K}(\theta|\theta^{(m)})
-
Q_n(\theta|\theta^{(m)})
\right|
\le
C_n\rho_{K-1}.
\]

Since

\[
\rho_{K-1}\to0
\qquad
(K\to\infty),
\]

we obtain

\[
\sup_{\theta\in\Theta}
\left|
Q_{n,K}(\theta|\theta^{(m)})
-
Q_n(\theta|\theta^{(m)})
\right|
\to0.
\]

This completes the proof.
\(\square\)

\end{proof}

\section{Approximate Monotonicity}

Define the observed-data log-likelihood

\[
\ell_n(\theta)
=
\sum_{i=1}^n
\log p_\theta(Y_i),
\]

where \(Y_i\) denotes the observed panel data for subject \(i\).

Let \(Z_i\) denote the corresponding complete latent trajectory. The EM auxiliary function is

\[
Q_n(\theta|\theta^{(m)})
=
\sum_{i=1}^n
E_{\theta^{(m)}}
\{
\log p_\theta(Y_i,Z_i)
\mid
Y_i
\}.
\]

For the exact EM algorithm, the standard Jensen inequality argument
implies

\[
\ell_n(\theta^{(m+1)})
\ge
\ell_n(\theta^{(m)}).
\]

Therefore, the exact EM algorithm is monotone.

\begin{theorem}[Approximate monotonicity under truncation]

Suppose the conditions of Theorem 2 hold. Let

\[
\epsilon_K
=
\sup_{\theta\in\Theta}
\left|
Q_{n,K}(\theta|\theta^{(m)})
-
Q_n(\theta|\theta^{(m)})
\right|.
\]

Then

\[
\ell_n(\theta^{(m+1)})
\ge
\ell_n(\theta^{(m)})
-
2\epsilon_K.
\]

Consequently, the deterministic uniformization EM algorithm is an asymptotically monotone generalized EM algorithm.
\end{theorem}

\begin{proof}
    Since the M-step maximizes the truncated auxiliary function,

\[
Q_{n,K}(\theta^{(m+1)}|\theta^{(m)})
\ge
Q_{n,K}(\theta^{(m)}|\theta^{(m)}).
\]

Using the uniform approximation bound,

\begin{align*}
Q_n(\theta^{(m+1)}|\theta^{(m)})
&\ge
Q_{n,K}(\theta^{(m+1)}|\theta^{(m)})
-
\epsilon_K
\\
&\ge
Q_{n,K}(\theta^{(m)}|\theta^{(m)})
-
\epsilon_K
\\
&\ge
Q_n(\theta^{(m)}|\theta^{(m)})
-
2\epsilon_K.
\end{align*}

By the standard EM minorization inequality,

\begin{align*}
\ell_n(\theta^{(m+1)})
-
\ell_n(\theta^{(m)})
\ge
Q_n(\theta^{(m+1)}|\theta^{(m)})
-
Q_n(\theta^{(m)}|\theta^{(m)}).
\end{align*}

Combining the above inequalities yields

\[
\ell_n(\theta^{(m+1)})
\ge
\ell_n(\theta^{(m)})
-
2\epsilon_K.
\]

\(\square\)

\end{proof}

\section{Asymptotic Theory}

\begin{lemma}[Observed likelihood approximation]
\label{lem:likelihood}
Under Assumption~\ref{ass:model},
let
\[
M_n(\theta)
=
\frac1n\ell_n(\theta),
\qquad
M_{n,K}(\theta)
=
\frac1n\ell_{n,K}(\theta).
\]

Then
\[
\sup_{\theta\in\Theta}
|M_{n,K}(\theta)-M_n(\theta)|
\le
C\rho_K.
\]

\end{lemma}

\begin{proof}

By Theorem~\ref{thm:transition},

\[
|P_{K,ab}(t;\theta)-P_{ab}(t;\theta)|
\le
\rho_K.
\]

By Assumption~\ref{ass:model},

\[
P_{ab}(t\mid x;\theta)\ge c_P>0.
\]

Hence the logarithm is Lipschitz continuous on
\([c_P,\infty)\).
Therefore,

\[
|\log P_{K,ab}-\log P_{ab}|
\le
C
|P_{K,ab}-P_{ab}|
\le
C\rho_K.
\]

Summing over all observation intervals yields

\[
|\ell_{n,K}(\theta)-\ell_n(\theta)|
\le
Cn\rho_K.
\]

Dividing by \(n\),

\[
|M_{n,K}(\theta)-M_n(\theta)|
\le
C\rho_K.
\]

Taking the supremum over \(\Theta\) completes the proof.

\end{proof}

\subsection{Consistency}

\begin{theorem}[Consistency]
\label{thm:consistency}

Suppose Assumptions
\ref{ass:compact}--\ref{ass:identifiability}
hold.

Let

\[
\hat\theta_{n,K_n}
=
\arg\max_{\theta\in\Theta}
M_{n,K_n}(\theta),
\]

where the truncation level \(K_n\) is chosen so that

\[
\rho_{K_n}\to0.
\]

Then

\[
\hat\theta_{n,K_n}
\overset{p}{\longrightarrow}
\theta_0.
\]

\end{theorem}
\begin{proof}

By Lemma~\ref{lem:likelihood},

\[
\sup_{\theta\in\Theta}
|M_{n,K_n}(\theta)-M_n(\theta)|
\to0.
\]

By Assumption~\ref{ass:ulln},

\[
\sup_{\theta\in\Theta}
|M_n(\theta)-M(\theta)|
\overset{p}{\to}
0.
\]

Therefore, by the triangle inequality,

\[
\sup_{\theta\in\Theta}
|M_{n,K_n}(\theta)-M(\theta)|
\overset{p}{\to}
0.
\]

Since \(M(\theta)\) is continuous on a compact parameter space and,
by Assumption~\ref{ass:identifiability}, has a unique maximizer at
\(\theta_0\), the Argmax Theorem implies

\[
\hat\theta_{n,K_n}
=
\arg\max_{\theta\in\Theta}
M_{n,K_n}(\theta)
\overset{p}{\longrightarrow}
\arg\max_{\theta\in\Theta}
M(\theta)
=
\theta_0.
\]

\end{proof}

\subsection{Score and Hessian approximation}

\paragraph{Definition of derivative truncation errors.}

The interval-level truncation errors for the transition probability, score function, and Hessian function were derived in Appendix~A of \citet{zhu2024uniformization}. For sample-level asymptotic analysis, Theorem~2 introduced the maximal interval-level transition-probability truncation error

\[
\rho_K
=
\max_{i,j}
\rho_K(\Delta t_{ij}).
\]

where $\rho_K(\Delta t_{ij})=1 -F_{\lambda_i\Delta t_{ij}}$ and \(F_{\lambda_i\Delta t_{ij}}\) denotes the cumulative distribution
function of a Poisson random variable with mean
\(\lambda_i\Delta t_{ij}\).

Similarly, define the maximal score-function truncation error

\[
\eta_K^{(1)}
=
\max_{i,j}
\left\{
2d_u\Delta t_{ij}
\left[
1-
F_{\lambda_i\Delta t_{ij}}(K-1)
\right]
\right\},
\]

and the maximal Hessian-function truncation error

\[
\eta_K^{(2)}
=
\max_{i,j}
\left\{
4d_ud_v\Delta t_{ij}^2
\left[
1-
F_{\lambda_i\Delta t_{ij}}(K-2)
\right]
+
d_{uv}\lambda_i\Delta t_{ij}
\left[
1-
F_{\lambda_i\Delta t_{ij}}(K-1)
\right]
\right\}.
\]

Finally, let

\[
\eta_K
=
\max
\left\{
\rho_K,
\eta_K^{(1)},
\eta_K^{(2)}
\right\}.
\]

Then

\[
\eta_K
\longrightarrow
0,
\qquad
K\to\infty,
\]

Thus, $\eta_K$ provides a common upper bound for the transition-probability, score-function, and Hessian-function truncation errors.

\begin{lemma}[Score and Hessian approximation]
\label{lem:score_hessian_approx}

Suppose Assumptions
\ref{ass:compact}--
\ref{ass:regularity}
hold. Let
\(\ell_n(\theta)\) denote the exact observed-data log-likelihood and
\(\ell_{n,K}(\theta)\) denote the deterministic uniformization truncated log-likelihood.
Then there exists a constant \(C<\infty\) such that
\[
\sup_{\theta\in\Theta}
\left\|
\frac{\partial}{\partial\theta}
\ell_{n,K}(\theta)
-
\frac{\partial}{\partial\theta}
\ell_n(\theta)
\right\|
\le
Cn\eta_K ,
\]
and
\[
\sup_{\theta\in\Theta}
\left\|
\frac{\partial^2}{\partial\theta\partial\theta^T}
\ell_{n,K}(\theta)
-
\frac{\partial^2}{\partial\theta\partial\theta^T}
\ell_n(\theta)
\right\|
\le
Cn\eta_K .
\]

Consequently, if \(K=K_n\) satisfies
\[
\sqrt n\,\eta_{K_n}\rightarrow0,
\]
then
\[
\frac1{\sqrt n}
\left[
\frac{\partial}{\partial\theta}
\ell_{n,K_n}(\theta_0)
-
\frac{\partial}{\partial\theta}
\ell_n(\theta_0)
\right]
=
o_p(1),
\]
and
\[
\sup_{\theta\in\Theta}
\left\|
\frac1n
\frac{\partial^2}{\partial\theta\partial\theta^T}
\ell_{n,K_n}(\theta)
-
\frac1n
\frac{\partial^2}{\partial\theta\partial\theta^T}
\ell_n(\theta)
\right\|
=
o_p(1).
\]

\end{lemma}

\begin{proof}

For observation interval \((i,j)\), define
\[
P_{ij}(\theta)
=
P_{Y_{i,j-1},Y_{ij}}
(\Delta t_{ij}\mid x_i;\theta),
\]
and
\[
P_{ij,K}(\theta)
=
P_{K,Y_{i,j-1},Y_{ij}}
(\Delta t_{ij}\mid x_i;\theta).
\]

The interval-level truncation results imply that, uniformly over \(\theta\in\Theta\),
\[
\left|
P_{ij,K}(\theta)-P_{ij}(\theta)
\right|
\le
C\rho_K
\le
C\eta_K.
\]
Similarly, the interval-level score-function truncation bound implies
\[
\left\|
\frac{\partial}{\partial\theta}
\log P_{ij,K}(\theta)
-
\frac{\partial}{\partial\theta}
\log P_{ij}(\theta)
\right\|
\le
C\eta_K^{(1)}
\le
C\eta_K.
\]
The interval-level Hessian-function truncation bound implies
\[
\left\|
\frac{\partial^2}{\partial\theta\partial\theta^T}
\log P_{ij,K}(\theta)
-
\frac{\partial^2}{\partial\theta\partial\theta^T}
\log P_{ij}(\theta)
\right\|
\le
C\eta_K^{(2)}
\le
C\eta_K.
\]

The observed-data score is the sum of interval-level scores:
\[
\frac{\partial}{\partial\theta}
\ell_n(\theta)
=
\sum_{i=1}^{n}
\sum_{j=1}^{m_i}
\frac{\partial}{\partial\theta}
\log P_{ij}(\theta),
\]
and
\[
\frac{\partial}{\partial\theta}
\ell_{n,K}(\theta)
=
\sum_{i=1}^{n}
\sum_{j=1}^{m_i}
\frac{\partial}{\partial\theta}
\log P_{ij,K}(\theta).
\]

Therefore,
\[
\left\|
\frac{\partial}{\partial\theta}
\ell_{n,K}(\theta)
-
\frac{\partial}{\partial\theta}
\ell_n(\theta)
\right\|
\le
\sum_{i=1}^{n}
\sum_{j=1}^{m_i}
C\eta_K.
\]

since \(m_i\le m_{\max}\), then
\[
\sum_{i=1}^{n}m_i
\le
nm_{\max}.
\]
Hence
\[
\sup_{\theta\in\Theta}
\left\|
\frac{\partial}{\partial\theta}
\ell_{n,K}(\theta)
-
\frac{\partial}{\partial\theta}
\ell_n(\theta)
\right\|
\le
Cn\eta_K.
\]

The same summation argument applies to the Hessian. Since the observed-data Hessian is the sum of interval-level Hessians,
\[
\frac{\partial^2}{\partial\theta\partial\theta^T}
\ell_n(\theta)
=
\sum_{i=1}^{n}
\sum_{j=1}^{m_i}
\frac{\partial^2}{\partial\theta\partial\theta^T}
\log P_{ij}(\theta),
\]
and
\[
\frac{\partial^2}{\partial\theta\partial\theta^T}
\ell_{n,K}(\theta)
=
\sum_{i=1}^{n}
\sum_{j=1}^{m_i}
\frac{\partial^2}{\partial\theta\partial\theta^T}
\log P_{ij,K}(\theta),
\]
we obtain
\[
\sup_{\theta\in\Theta}
\left\|
\frac{\partial^2}{\partial\theta\partial\theta^T}
\ell_{n,K}(\theta)
-
\frac{\partial^2}{\partial\theta\partial\theta^T}
\ell_n(\theta)
\right\|
\le
Cn\eta_K.
\]

If \(\sqrt n\,\eta_{K_n}\to0\), then
\[
\frac1{\sqrt n}
\left\|
\frac{\partial}{\partial\theta}
\ell_{n,K_n}(\theta_0)
-
\frac{\partial}{\partial\theta}
\ell_n(\theta_0)
\right\|
\le
C\sqrt n\,\eta_{K_n}
=
o(1).
\]
Therefore,
\[
\frac1{\sqrt n}
\left[
\frac{\partial}{\partial\theta}
\ell_{n,K_n}(\theta_0)
-
\frac{\partial}{\partial\theta}
\ell_n(\theta_0)
\right]
=
o_p(1).
\]

For the Hessian,
\[
\sup_{\theta\in\Theta}
\left\|
\frac1n
\frac{\partial^2}{\partial\theta\partial\theta^T}
\ell_{n,K_n}(\theta)
-
\frac1n
\frac{\partial^2}{\partial\theta\partial\theta^T}
\ell_n(\theta)
\right\|
\le
C\eta_{K_n}
=
o(1).
\]
Thus,
\[
\sup_{\theta\in\Theta}
\left\|
\frac1n
\frac{\partial^2}{\partial\theta\partial\theta^T}
\ell_{n,K_n}(\theta)
-
\frac1n
\frac{\partial^2}{\partial\theta\partial\theta^T}
\ell_n(\theta)
\right\|
=
o_p(1).
\]

This proves the score and Hessian approximation results.

\end{proof}

\subsection{Asymptotic normality}

\begin{theorem}[Asymptotic normality]
\label{thm:asymptotic_normality}

Suppose Assumptions 1-6 hold.

Assume that
\[
\sqrt n\,\eta_{K_n}
\longrightarrow
0,
\]
where \(\eta_{K_n}\) is the uniform interval-level truncation error in Lemma~\ref{lem:score_hessian_approx}.

Then
\[
\sqrt n
\left(
\hat\theta_{n,K_n}
-
\theta_0
\right)
\Rightarrow
N
\left(
0,
I(\theta_0)^{-1}
\right).
\]

\end{theorem}

\begin{proof}

By Theorem~\ref{thm:consistency},
\[
\hat\theta_{n,K_n}
\overset{p}{\longrightarrow}
\theta_0.
\]

Since \(\theta_0\) lies in the interior of \(\Theta\), it follows that
\(\hat\theta_{n,K_n}\) is an interior maximizer with probability tending
to one. Hence,

\[
\frac{\partial}{\partial\theta}
\ell_{n,K_n}
(\hat\theta_{n,K_n})
=
0
\]

with probability tending to one.
Therefore, applying a Taylor expansion around \(\theta_0\),
\[
0
=
\frac{\partial}{\partial\theta}
\ell_{n,K_n}(\theta_0)
+
\frac{\partial^2}{\partial\theta\partial\theta^T}
\ell_{n,K_n}(\tilde\theta_n)
\left(
\hat\theta_{n,K_n}
-
\theta_0
\right),
\]
where \(\tilde\theta_n\) lies between
\(\hat\theta_{n,K_n}\) and \(\theta_0\). Moreover, because
\(\tilde\theta_n\) lies between two sequences converging to
\(\theta_0\),

\[
\tilde\theta_n
\overset{p}{\longrightarrow}
\theta_0.
\]

Rearranging gives
\[
\sqrt n
\left(
\hat\theta_{n,K_n}
-
\theta_0
\right)
=
-
\left[
\frac1n
\frac{\partial^2}{\partial\theta\partial\theta^T}
\ell_{n,K_n}(\tilde\theta_n)
\right]^{-1}
\left[
\frac1{\sqrt n}
\frac{\partial}{\partial\theta}
\ell_{n,K_n}(\theta_0)
\right].
\]

By Lemma~\ref{lem:score_hessian_approx},
\[
\frac1{\sqrt n}
\frac{\partial}{\partial\theta}
\ell_{n,K_n}(\theta_0)
=
\frac1{\sqrt n}
\frac{\partial}{\partial\theta}
\ell_n(\theta_0)
+
o_p(1),
\]
because \(\sqrt n\,\eta_{K_n}\to0\). Also,
\[
\frac1n
\frac{\partial^2}{\partial\theta\partial\theta^T}
\ell_{n,K_n}(\tilde\theta_n)
=
\frac1n
\frac{\partial^2}{\partial\theta\partial\theta^T}
\ell_n(\tilde\theta_n)
+
o_p(1).
\]

For the exact likelihood, standard likelihood theory gives
\[
\frac1{\sqrt n}
\frac{\partial}{\partial\theta}
\ell_n(\theta_0)
\Rightarrow
N
\left(
0,
I(\theta_0)
\right).
\]
Moreover, since \(\tilde\theta_n\overset{p}{\to}\theta_0\) and the Hessian convergence holds uniformly in a neighborhood of \(\theta_0\),
\[
-
\frac1n
\frac{\partial^2}{\partial\theta\partial\theta^T}
\ell_n(\tilde\theta_n)
\overset{p}{\longrightarrow}
I(\theta_0).
\]

Therefore,
\[
\frac1{\sqrt n}
\frac{\partial}{\partial\theta}
\ell_{n,K_n}(\theta_0)
\Rightarrow
N
\left(
0,
I(\theta_0)
\right),
\]
and
\[
-
\frac1n
\frac{\partial^2}{\partial\theta\partial\theta^T}
\ell_{n,K_n}(\tilde\theta_n)
\overset{p}{\longrightarrow}
I(\theta_0).
\]

Applying Slutsky's theorem yields
\[
\sqrt n
\left(
\hat\theta_{n,K_n}
-
\theta_0
\right)
\Rightarrow
N
\left(
0,
I(\theta_0)^{-1}
\right).
\]

\end{proof}

\section{Additional Simulation Results}

Coefficient-specific results from the new 100-replicate, $N=500$ experiment
are reported below. Both methods were applied to every scenario. The deterministic EM standard
errors was estimated by Louis' observed information. One Scenario B
\texttt{msm} replicate produced point estimates but no standard errors because
its numerical Hessian was not positive definite; the corresponding coverage
entries are averaged over available standard errors.

\begin{table}[!ht]
\centering
\scriptsize
\caption{Coefficient estimation results for Scenario A.}
\label{tab:supp_coef_A}
\resizebox{\textwidth}{!}{%
\begin{tabular}{llrccc|ccc}
\toprule
&&& \multicolumn{3}{c}{Deterministic Uniformization EM} & \multicolumn{3}{c}{R \texttt{msm}} \\
\cmidrule(lr){4-6}\cmidrule(lr){7-9}
Transition & Coefficient & True & Mean est. & Mean SE & Coverage & Mean est. & Mean SE & Coverage \\
\midrule
$1\to2$ & Intercept & -1.800 & -1.776 & 0.109 & 0.940 & -1.776 & 0.109 & 0.940 \\
 & $x_1$ & 0.450 & 0.449 & 0.081 & 0.980 & 0.449 & 0.081 & 0.980 \\
 & $x_2$ & -0.200 & -0.212 & 0.158 & 0.960 & -0.212 & 0.158 & 0.960 \\
\addlinespace
$1\to3$ & Intercept & -2.100 & -2.113 & 0.134 & 0.940 & -2.113 & 0.134 & 0.940 \\
 & $x_1$ & -0.250 & -0.251 & 0.088 & 0.970 & -0.251 & 0.088 & 0.970 \\
 & $x_2$ & 0.300 & 0.315 & 0.167 & 0.970 & 0.314 & 0.167 & 0.970 \\
\addlinespace
$2\to3$ & Intercept & -1.600 & -1.636 & 0.192 & 0.940 & -1.636 & 0.192 & 0.940 \\
 & $x_1$ & 0.350 & 0.368 & 0.138 & 0.930 & 0.368 & 0.138 & 0.930 \\
 & $x_2$ & -0.300 & -0.270 & 0.272 & 0.910 & -0.270 & 0.272 & 0.910 \\
\bottomrule
\end{tabular}%
}
\end{table}

\begin{table}[!ht]
\centering
\scriptsize
\caption{Coefficient estimation results for Scenario B.}
\label{tab:supp_coef_B}
\resizebox{\textwidth}{!}{%
\begin{tabular}{llrccc|ccc}
\toprule
&&& \multicolumn{3}{c}{Deterministic Uniformization EM} & \multicolumn{3}{c}{R \texttt{msm}} \\
\cmidrule(lr){4-6}\cmidrule(lr){7-9}
Transition & Coefficient & True & Mean est. & Mean SE & Coverage & Mean est. & Mean SE & Coverage \\
\midrule
$1\to2$ & Intercept & -1.800 & -1.793 & 0.091 & 0.970 & -1.793 & 0.091 & 0.970 \\
 & $x_1$ & 0.450 & 0.461 & 0.067 & 0.960 & 0.461 & 0.067 & 0.960 \\
 & $x_2$ & -0.200 & -0.219 & 0.130 & 0.970 & -0.219 & 0.130 & 0.970 \\
\addlinespace
$2\to3$ & Intercept & -1.600 & -1.599 & 0.153 & 0.920 & -1.599 & 0.153 & 0.919 \\
 & $x_1$ & 0.350 & 0.352 & 0.114 & 0.930 & 0.352 & 0.114 & 0.929 \\
 & $x_2$ & -0.300 & -0.339 & 0.221 & 0.920 & -0.339 & 0.221 & 0.919 \\
\addlinespace
$3\to4$ & Intercept & -1.700 & -1.740 & 0.314 & 0.960 & -1.739 & 0.314 & 0.960 \\
 & $x_1$ & -0.250 & -0.292 & 0.230 & 0.960 & -0.292 & 0.230 & 0.960 \\
 & $x_2$ & 0.300 & 0.334 & 0.447 & 0.930 & 0.333 & 0.447 & 0.929 \\
\bottomrule
\end{tabular}%
}
\end{table}

\begin{table}[!ht]
\centering
\scriptsize
\caption{Coefficient estimation results for Scenario C-low.}
\label{tab:supp_coef_C_low}
\resizebox{\textwidth}{!}{%
\begin{tabular}{llrccc|ccc}
\toprule
&&& \multicolumn{3}{c}{Deterministic Uniformization EM} & \multicolumn{3}{c}{R \texttt{msm}} \\
\cmidrule(lr){4-6}\cmidrule(lr){7-9}
Transition & Coefficient & True & Mean est. & Mean SE & Coverage & Mean est. & Mean SE & Coverage \\
\midrule
$1\to2$ & Intercept & -0.076 & -0.075 & 0.072 & 0.930 & -0.075 & 0.072 & 0.930 \\
 & $x_1$ & 0.400 & 0.403 & 0.052 & 0.950 & 0.403 & 0.052 & 0.950 \\
 & $x_2$ & 0.050 & 0.061 & 0.101 & 0.910 & 0.061 & 0.101 & 0.910 \\
\addlinespace
$1\to3$ & Intercept & -2.871 & -2.879 & 0.304 & 0.960 & -2.879 & 0.304 & 0.960 \\
 & $x_1$ & 0.000 & -0.015 & 0.225 & 0.990 & -0.014 & 0.225 & 0.990 \\
 & $x_2$ & -0.250 & -0.315 & 0.452 & 0.930 & -0.313 & 0.452 & 0.930 \\
\addlinespace
$2\to4$ & Intercept & -1.550 & -1.551 & 0.097 & 0.930 & -1.551 & 0.097 & 0.930 \\
 & $x_1$ & 0.200 & 0.204 & 0.071 & 0.940 & 0.204 & 0.071 & 0.940 \\
 & $x_2$ & -0.100 & -0.114 & 0.137 & 0.950 & -0.115 & 0.137 & 0.950 \\
\addlinespace
$3\to4$ & Intercept & -1.550 & -1.669 & 0.611 & 0.940 & -1.659 & 0.594 & 0.940 \\
 & $x_1$ & 0.200 & 0.208 & 0.434 & 0.890 & 0.214 & 0.414 & 0.880 \\
 & $x_2$ & -0.100 & -0.347 & 1.455 & 0.960 & -0.248 & 1.070 & 0.960 \\
\bottomrule
\end{tabular}%
}
\end{table}

\begin{table}[!ht]
\centering
\scriptsize
\caption{Coefficient estimation results for Scenario C-moderate.}
\label{tab:supp_coef_C_moderate}
\resizebox{\textwidth}{!}{%
\begin{tabular}{llrccc|ccc}
\toprule
&&& \multicolumn{3}{c}{Deterministic Uniformization EM} & \multicolumn{3}{c}{R \texttt{msm}} \\
\cmidrule(lr){4-6}\cmidrule(lr){7-9}
Transition & Coefficient & True & Mean est. & Mean SE & Coverage & Mean est. & Mean SE & Coverage \\
\midrule
$1\to2$ & Intercept & -0.382 & -0.378 & 0.086 & 0.960 & -0.378 & 0.086 & 0.960 \\
 & $x_1$ & 0.400 & 0.404 & 0.061 & 0.920 & 0.404 & 0.061 & 0.920 \\
 & $x_2$ & 0.050 & 0.041 & 0.118 & 0.970 & 0.041 & 0.118 & 0.970 \\
\addlinespace
$1\to3$ & Intercept & -1.079 & -1.093 & 0.123 & 0.910 & -1.093 & 0.123 & 0.910 \\
 & $x_1$ & 0.000 & 0.001 & 0.092 & 0.950 & 0.001 & 0.092 & 0.950 \\
 & $x_2$ & -0.250 & -0.264 & 0.179 & 0.950 & -0.264 & 0.179 & 0.950 \\
\addlinespace
$2\to4$ & Intercept & -1.550 & -1.567 & 0.120 & 0.990 & -1.567 & 0.120 & 0.990 \\
 & $x_1$ & 0.200 & 0.193 & 0.085 & 0.930 & 0.193 & 0.085 & 0.930 \\
 & $x_2$ & -0.100 & -0.100 & 0.164 & 0.940 & -0.100 & 0.164 & 0.940 \\
\addlinespace
$3\to4$ & Intercept & -1.550 & -1.548 & 0.180 & 0.930 & -1.548 & 0.180 & 0.930 \\
 & $x_1$ & 0.200 & 0.221 & 0.142 & 0.920 & 0.221 & 0.142 & 0.920 \\
 & $x_2$ & -0.100 & -0.120 & 0.276 & 0.980 & -0.119 & 0.276 & 0.980 \\
\bottomrule
\end{tabular}%
}
\end{table}

\begin{table}[!ht]
\centering
\scriptsize
\caption{Coefficient estimation results for Scenario C-high.}
\label{tab:supp_coef_C_high}
\resizebox{\textwidth}{!}{%
\begin{tabular}{llrccc|ccc}
\toprule
&&& \multicolumn{3}{c}{Deterministic Uniformization EM} & \multicolumn{3}{c}{R \texttt{msm}} \\
\cmidrule(lr){4-6}\cmidrule(lr){7-9}
Transition & Coefficient & True & Mean est. & Mean SE & Coverage & Mean est. & Mean SE & Coverage \\
\midrule
$1\to2$ & Intercept & -0.718 & -0.698 & 0.103 & 0.920 & -0.698 & 0.103 & 0.920 \\
 & $x_1$ & 0.400 & 0.394 & 0.072 & 0.950 & 0.394 & 0.072 & 0.950 \\
 & $x_2$ & 0.050 & 0.028 & 0.139 & 0.920 & 0.028 & 0.139 & 0.920 \\
\addlinespace
$1\to3$ & Intercept & -0.568 & -0.552 & 0.096 & 0.940 & -0.552 & 0.096 & 0.940 \\
 & $x_1$ & 0.000 & 0.001 & 0.071 & 0.940 & 0.001 & 0.071 & 0.940 \\
 & $x_2$ & -0.250 & -0.284 & 0.139 & 0.980 & -0.283 & 0.139 & 0.980 \\
\addlinespace
$2\to4$ & Intercept & -1.550 & -1.542 & 0.146 & 0.960 & -1.542 & 0.146 & 0.960 \\
 & $x_1$ & 0.200 & 0.181 & 0.102 & 0.960 & 0.181 & 0.102 & 0.960 \\
 & $x_2$ & -0.100 & -0.111 & 0.197 & 0.960 & -0.111 & 0.197 & 0.960 \\
\addlinespace
$3\to4$ & Intercept & -1.550 & -1.549 & 0.135 & 0.950 & -1.549 & 0.135 & 0.950 \\
 & $x_1$ & 0.200 & 0.222 & 0.105 & 0.930 & 0.222 & 0.105 & 0.930 \\
 & $x_2$ & -0.100 & -0.099 & 0.206 & 0.950 & -0.099 & 0.206 & 0.950 \\
\bottomrule
\end{tabular}%
}
\end{table}

\begin{table}[!ht]
\centering
\scriptsize
\caption{Coefficient estimation results for Scenario D.}
\label{tab:supp_coef_D}
\resizebox{\textwidth}{!}{%
\begin{tabular}{llrccc|ccc}
\toprule
&&& \multicolumn{3}{c}{Deterministic Uniformization EM} & \multicolumn{3}{c}{R \texttt{msm}} \\
\cmidrule(lr){4-6}\cmidrule(lr){7-9}
Transition & Coefficient & True & Mean est. & Mean SE & Coverage & Mean est. & Mean SE & Coverage \\
\midrule
$1\to2$ & Intercept & -0.693 & -0.706 & 0.104 & 0.930 & -0.706 & 0.104 & 0.930 \\
 & $x_1$ & 0.400 & 0.404 & 0.080 & 0.950 & 0.404 & 0.080 & 0.950 \\
 & $x_2$ & -0.200 & -0.176 & 0.151 & 0.920 & -0.176 & 0.151 & 0.920 \\
\addlinespace
$1\to3$ & Intercept & -0.693 & -0.690 & 0.097 & 0.950 & -0.690 & 0.097 & 0.950 \\
 & $x_1$ & -0.350 & -0.346 & 0.068 & 0.910 & -0.346 & 0.068 & 0.910 \\
 & $x_2$ & 0.250 & 0.260 & 0.129 & 0.930 & 0.260 & 0.129 & 0.930 \\
\addlinespace
$2\to1$ & Intercept & -2.800 & -2.895 & 0.369 & 0.960 & -2.894 & 0.369 & 0.960 \\
 & $x_1$ & -0.250 & -0.300 & 0.267 & 0.930 & -0.300 & 0.266 & 0.930 \\
 & $x_2$ & 0.200 & 0.276 & 0.500 & 0.990 & 0.281 & 0.499 & 0.980 \\
\addlinespace
$2\to4$ & Intercept & -1.600 & -1.602 & 0.155 & 0.940 & -1.602 & 0.155 & 0.940 \\
 & $x_1$ & 0.300 & 0.308 & 0.117 & 0.940 & 0.308 & 0.117 & 0.940 \\
 & $x_2$ & -0.250 & -0.256 & 0.222 & 0.990 & -0.256 & 0.222 & 0.990 \\
\addlinespace
$3\to4$ & Intercept & -1.600 & -1.606 & 0.141 & 0.950 & -1.606 & 0.141 & 0.950 \\
 & $x_1$ & -0.200 & -0.209 & 0.090 & 0.960 & -0.209 & 0.090 & 0.960 \\
 & $x_2$ & 0.300 & 0.287 & 0.173 & 0.930 & 0.287 & 0.173 & 0.930 \\
\bottomrule
\end{tabular}%
}
\end{table}

\clearpage
\section{Implementation Details}

\subsection{Choice of Truncation Level}

The infinite Poisson summation was truncated at the smallest integer
\(K\) satisfying

\[
P(N>K)<10^{-10},
\qquad
N\sim\mathrm{Poisson}(\lambda t).
\]

\subsection{Convergence Criterion}

The EM algorithm terminated when

\[
\max_{r,s}
\|
\beta_{rs}^{(m+1)}
-
\beta_{rs}^{(m)}
\|_\infty
<
10^{-6}.
\]

\subsection{Louis Observed Information and Its Joint Reward Recursion}
\label{sec:supp_louis}

This section describes the deterministic calculation of the observed
information matrix used in the implementation. The main idea is to
express the complete-data score as a combination of jump and occupation
rewards under uniformization, and then propagate their conditional first
and second moments through a single forward recursion.

Let $h$ index an observation interval of length $t_h$, with observed
initial and terminal states $a$ and $b$, respectively. Let $x_h$ denote
the covariate vector, assumed fixed at its interval-start value.
Stack the transition-specific regression coefficients as
\[
\theta=
(\beta_{r_1s_1}^\top,\ldots,\beta_{r_Es_E}^\top)^\top.
\]
For each allowed transition $r\to s$, define $e_{rs}(x_h)$ as the
vector containing $x_h$ in the coefficient block corresponding to
$r\to s$ and zeros elsewhere.

\subsubsection{Complete-data score and Louis' identity}

If the complete continuous-time trajectory were observed, the
transition counts $N_{rs,h}$ and state occupation times $T_{r,h}$
would be available. Under the log-linear transition intensity model
\[
q_{rs,h}=\exp\{x_h^\top\beta_{rs}\},
\]
the complete-data score for interval $h$ is
\begin{equation}
S_{c,h}(\theta)
=
\sum_{r\ne s}e_{rs}(x_h)N_{rs,h}
-
\sum_{r\ne s}q_{rs,h}e_{rs}(x_h)T_{r,h},
\label{eq:supp_complete_score}
\end{equation}
where the sums are over allowed transitions.

The corresponding complete-data Hessian is block diagonal, with
\begin{equation}
H_{c,h}^{(rs,rs)}(\theta)
=
-q_{rs,h}T_{r,h}x_hx_h^\top,
\qquad
H_{c,h}^{(rs,uv)}(\theta)=0
\quad (rs\ne uv).
\label{eq:supp_complete_hessian}
\end{equation}

Because the intermediate states and transition times are unobserved,
the complete-data score and Hessian must be averaged over the
conditional distribution of the latent trajectory.

Louis' identity expresses the interval-specific observed information
as
\begin{equation}
\begin{aligned}
I_{\mathrm{obs},h}(\theta)
={}&-E(H_{c,h}\mid\mathcal O_h)\\
&-\Big\{
E(S_{c,h}S_{c,h}^\top\mid\mathcal O_h)
-
E(S_{c,h}\mid\mathcal O_h)
E(S_{c,h}\mid\mathcal O_h)^\top
\Big\},
\end{aligned}
\label{eq:supp_louis}
\end{equation}
where $\mathcal O_h$ denotes the observed states at the interval
boundaries.

Thus, in addition to the expected complete-data Hessian, evaluation
of the observed information requires two conditional score moments:
\[
E(S_{c,h}\mid\mathcal O_h),
\qquad
E(S_{c,h}S_{c,h}^\top\mid\mathcal O_h).
\]
The second moment is particularly important because it includes
cross-products between jump and occupation contributions, as well
as the uncertainty associated with unobserved transition times.

Conditional on the observed states at interval boundaries, latent
bridges in distinct intervals are independent under the assumed
Markov model. Consequently, the sample observed information is
obtained by summing the interval contributions,
\[
I_{\mathrm{obs}}(\theta)
=
\sum_h I_{\mathrm{obs},h}(\theta).
\]

\subsubsection{Jump and occupation rewards}

We first represent the complete-data score in terms of additive
rewards along a uniformized trajectory.

For an interval of length $t_h$, let $K$ denote the number of
uniformization events and let
\[
Z_0,Z_1,\ldots,Z_k
\]
be the embedded Markov path conditional on $K=k$.
The embedded transition matrix is
\[
R=I+\frac{Q_h}{\lambda},
\]
where $\lambda$ is a valid uniformization rate for the interval.

Define the jump reward associated with an embedded transition
$i\to j$ by
\begin{equation}
a(i,j)=
\begin{cases}
e_{ij}(x_h),
& i\ne j \text{ and } i\to j \text{ is allowed},\\
0,
& \text{otherwise},
\end{cases}
\label{eq:supp_jump_reward}
\end{equation}
and define the occupation reward associated with state $i$ by
\begin{equation}
g(i)
=
-\sum_{s:i\to s}q_{is,h}e_{is}(x_h).
\label{eq:supp_occupation_reward}
\end{equation}

The jump reward records the contribution of each genuine state
transition to the complete-data score. Self-transitions introduced
by uniformization contribute zero jump reward. The occupation
reward represents the score contribution per unit time spent in
a given state.

Let $\Delta_0,\ldots,\Delta_k$ denote the spacings between
successive uniformization events, including the initial and
terminal segments. Conditional on the embedded path and its
spacings, the complete-data score can be written as
\begin{equation}
S_{c,h}
=
\sum_{u=1}^k a(Z_{u-1},Z_u)
+
\sum_{u=0}^k\Delta_u g(Z_u).
\label{eq:supp_score_reward_representation}
\end{equation}

Define the accumulated jump and occupation rewards
\begin{equation}
\begin{aligned}
J_k
&=\sum_{u=1}^k a(Z_{u-1},Z_u),\\
G_k
&=\sum_{u=0}^k g(Z_u),\\
D_k
&=\sum_{u=0}^k g(Z_u)g(Z_u)^\top.
\end{aligned}
\label{eq:supp_rewards}
\end{equation}

The quantity $J_k$ is the accumulated jump contribution,
whereas $G_k$ is the sum of the occupation reward vectors over
the embedded segments. The matrix $D_k$ collects their individual
outer products and will account for the additional variation
introduced by the random spacings.

Conditional on $K=k$, the $L=k+1$ spacings follow a scaled
Dirichlet$(1,\ldots,1)$ distribution:
\[
(\Delta_0,\ldots,\Delta_k)
\sim
t_h\,\operatorname{Dirichlet}(1,\ldots,1).
\]
In particular,
\[
E(\Delta_u\mid K=k)=\frac{t_h}{L}.
\]
Writing $c_k=t_h/L$, integration over the spacings gives
\begin{equation}
E(S_{c,h}\mid Z_0,\ldots,Z_k,K=k)
=
J_k+c_kG_k.
\label{eq:supp_score_first_path}
\end{equation}

For the second moment, the Dirichlet spacing covariance must
also be included. Specifically,
\[
\operatorname{Var}(\Delta_u\mid K=k)
=
\frac{t_h^2(L-1)}{L^2(L+1)}
\]
and, for $u\ne v$,
\[
\operatorname{Cov}(\Delta_u,\Delta_v\mid K=k)
=
-\frac{t_h^2}{L^2(L+1)}.
\]

It follows that
\begin{equation}
\begin{aligned}
&\operatorname{Var}
\left(
\sum_{u=0}^k\Delta_u g(Z_u)
\;\middle|\;
Z_0,\ldots,Z_k,K=k
\right)\\
&\qquad=
\frac{t_h^2}{L^2(L+1)}
\left\{
LD_k-G_kG_k^\top
\right\}.
\end{aligned}
\label{eq:supp_dirichlet_reward_variance}
\end{equation}

Combining this variance with the conditional mean yields
\begin{align}
&E(S_{c,h}S_{c,h}^\top
\mid Z_0,\ldots,Z_k,K=k)
\nonumber\\
&\quad=
(J_k+c_kG_k)(J_k+c_kG_k)^\top
+
\frac{t_h^2}{L^2(L+1)}
\{LD_k-G_kG_k^\top\}.
\label{eq:supp_score_second_path}
\end{align}

Expanding the first term gives
\begin{align}
&(J_k+c_kG_k)(J_k+c_kG_k)^\top
\nonumber\\
&\quad=
J_kJ_k^\top
+c_kJ_kG_k^\top
+c_kG_kJ_k^\top
+c_k^2G_kG_k^\top.
\label{eq:supp_score_product_expansion}
\end{align}

Equations~\eqref{eq:supp_score_first_path}--%
\eqref{eq:supp_score_product_expansion} identify the quantities
that must be averaged over the embedded Markov paths.
The first score moment requires $J_k$ and $G_k$, whereas
the second score moment requires
\[
J_kJ_k^\top,\qquad
J_kG_k^\top,\qquad
G_kG_k^\top,\qquad
D_k.
\]
We next show that all these quantities can be propagated jointly
without enumerating individual embedded paths.

\subsubsection{One-pass forward moment recursion}

\paragraph{Additive updates along the embedded path.}

Consider an embedded path $Z_0,\ldots,Z_k$ ending in state $i$,
followed by an additional transition $i\to j$.
Write
\[
a_{ij}=a(i,j),
\qquad
g_j=g(j).
\]
By their definitions, the accumulated rewards satisfy
\begin{equation}
\begin{aligned}
J_{k+1}&=J_k+a_{ij},\\
G_{k+1}&=G_k+g_j,\\
D_{k+1}&=D_k+g_jg_j^\top.
\end{aligned}
\label{eq:supp_reward_updates}
\end{equation}

These additive updates are the basis of the forward recursion.
In particular, products of the updated rewards can be expanded
in terms of their previous first and second moments, so no
higher-order reward quantities are required.

\paragraph{Endpoint-specific weighted moments.}

Let $\mathcal P_k(i)$ denote the set of embedded paths
\[
\omega=(z_0,\ldots,z_k)
\]
that start at the observed initial state $a$ and end at state
$i$ after $k$ embedded transitions.

The probability of a path under the embedded chain is
\[
\pi(\omega)
=
\prod_{u=1}^kR_{z_{u-1},z_u}.
\]

We first define the total probability of reaching state $i$
after $k$ steps:
\begin{equation}
w_k(i)
=
\sum_{\omega\in\mathcal P_k(i)}
\pi(\omega)
=
(R^k)_{ai}.
\label{eq:supp_endpoint_probability}
\end{equation}

The corresponding unnormalized first reward moments are
\begin{equation}
\begin{aligned}
m_k^J(i)
&=
\sum_{\omega\in\mathcal P_k(i)}
\pi(\omega)J_k(\omega),\\
m_k^G(i)
&=
\sum_{\omega\in\mathcal P_k(i)}
\pi(\omega)G_k(\omega).
\end{aligned}
\label{eq:supp_first_reward_moments}
\end{equation}

These quantities are probability-weighted reward sums rather
than normalized conditional expectations. For example, whenever
$w_k(i)>0$,
\[
E(J_k\mid Z_0=a,Z_k=i,K=k)
=
\frac{m_k^J(i)}{w_k(i)}.
\]

Similarly, define the unnormalized second reward moments
\begin{equation}
\begin{aligned}
M_k^{JJ}(i)
&=
\sum_{\omega\in\mathcal P_k(i)}
\pi(\omega)J_k(\omega)J_k(\omega)^\top,\\
M_k^{JG}(i)
&=
\sum_{\omega\in\mathcal P_k(i)}
\pi(\omega)J_k(\omega)G_k(\omega)^\top,\\
M_k^{GG}(i)
&=
\sum_{\omega\in\mathcal P_k(i)}
\pi(\omega)G_k(\omega)G_k(\omega)^\top,\\
M_k^D(i)
&=
\sum_{\omega\in\mathcal P_k(i)}
\pi(\omega)D_k(\omega).
\end{aligned}
\label{eq:supp_second_reward_moments}
\end{equation}

The path sums above define the required moments; they are
not evaluated by explicit path enumeration. Instead, we
derive recursions that update these moments directly from
their values at the preceding step.

\paragraph{Endpoint probability and first-moment recursions.}

Every path ending in state $j$ after $k+1$ steps can be
uniquely decomposed into a path ending in some state $i$
after $k$ steps, followed by one additional transition
$i\to j$.

If $\omega\in\mathcal P_k(i)$, the extended path has
probability $\pi(\omega)R_{ij}$. Therefore,
\begin{equation}
\begin{aligned}
w_{k+1}(j)
&=
\sum_i
\sum_{\omega\in\mathcal P_k(i)}
\pi(\omega)R_{ij}\\
&=
\sum_iR_{ij}w_k(i).
\end{aligned}
\label{eq:supp_probability_recursion}
\end{equation}

For the jump reward, the additive update
$J_{k+1}=J_k+a_{ij}$ gives
\begin{align}
m_{k+1}^J(j)
={}&
\sum_i
\sum_{\omega\in\mathcal P_k(i)}
\pi(\omega)R_{ij}
\{J_k(\omega)+a_{ij}\}
\nonumber\\
={}&
\sum_iR_{ij}
\left\{
\sum_{\omega\in\mathcal P_k(i)}
\pi(\omega)J_k(\omega)
+
\sum_{\omega\in\mathcal P_k(i)}
\pi(\omega)a_{ij}
\right\}
\nonumber\\
={}&
\sum_iR_{ij}
\{m_k^J(i)+w_k(i)a_{ij}\}.
\label{eq:supp_jump_first_recursion}
\end{align}

The last equality follows because $a_{ij}$ depends only
on the newly added transition and is therefore constant
across all historical paths ending in $i$.

Applying the same argument to
$G_{k+1}=G_k+g_j$ yields
\begin{equation}
m_{k+1}^G(j)
=
\sum_iR_{ij}
\{m_k^G(i)+w_k(i)g_j\}.
\label{eq:supp_occupation_first_recursion}
\end{equation}

Thus, the first reward moments at step $k+1$ depend
only on the endpoint probabilities and first moments
already available at step $k$.

\paragraph{Second-moment recursions.}

The second moments follow from the same path-extension
argument, now applied to products of the accumulated
rewards.

We first consider the jump--occupation cross moment,
which is needed to account for the dependence between
the jump and occupation contributions to the score.

By definition,
\begin{align}
M_{k+1}^{JG}(j)
={}&
\sum_i
\sum_{\omega\in\mathcal P_k(i)}
\pi(\omega)R_{ij}
\nonumber\\
&\quad\times
\{J_k(\omega)+a_{ij}\}
\{G_k(\omega)+g_j\}^\top.
\label{eq:supp_JG_path_sum}
\end{align}

For each extended path, the reward product expands as
\begin{align}
&(J_k+a_{ij})(G_k+g_j)^\top
\nonumber\\
&\quad=
J_kG_k^\top
+J_kg_j^\top
+a_{ij}G_k^\top
+a_{ij}g_j^\top.
\label{eq:supp_JG_product_expansion}
\end{align}

Substituting this expansion into
Equation~\eqref{eq:supp_JG_path_sum} and grouping the
historical path sums gives
\begin{align}
M_{k+1}^{JG}(j)
={}&
\sum_iR_{ij}
\Bigg\{
\underbrace{
\sum_{\omega\in\mathcal P_k(i)}
\pi(\omega)J_kG_k^\top
}_{M_k^{JG}(i)}
\nonumber\\
&\quad+
\underbrace{
\sum_{\omega\in\mathcal P_k(i)}
\pi(\omega)J_k
}_{m_k^J(i)}
g_j^\top
\nonumber\\
&\quad+
a_{ij}
\underbrace{
\sum_{\omega\in\mathcal P_k(i)}
\pi(\omega)G_k^\top
}_{m_k^G(i)^\top}
\nonumber\\
&\quad+
\underbrace{
\sum_{\omega\in\mathcal P_k(i)}
\pi(\omega)
}_{w_k(i)}
a_{ij}g_j^\top
\Bigg\}.
\label{eq:supp_JG_weighted_expansion}
\end{align}

Consequently,
\begin{equation}
\begin{aligned}
M_{k+1}^{JG}(j)
=
\sum_iR_{ij}\big\{
&M_k^{JG}(i)
+m_k^J(i)g_j^\top\\
&+a_{ij}m_k^G(i)^\top
+w_k(i)a_{ij}g_j^\top
\big\}.
\end{aligned}
\label{eq:supp_JG_recursion}
\end{equation}

The four terms represent the previous jump--occupation
product, the interaction between the previous jump
reward and the new occupation reward, the interaction
between the new jump reward and the previous occupation
reward, and the product of the two newly added rewards.

The remaining second-moment recursions follow from
the analogous product expansions
\begin{align*}
&(J_k+a_{ij})(J_k+a_{ij})^\top\\
&\quad=
J_kJ_k^\top
+J_ka_{ij}^\top
+a_{ij}J_k^\top
+a_{ij}a_{ij}^\top,
\end{align*}
and
\begin{align*}
&(G_k+g_j)(G_k+g_j)^\top\\
&\quad=
G_kG_k^\top
+G_kg_j^\top
+g_jG_k^\top
+g_jg_j^\top.
\end{align*}

After probability-weighted summation over the historical
paths, these give
\begin{align}
M_{k+1}^{JJ}(j)
={}&
\sum_iR_{ij}
\big\{
M_k^{JJ}(i)
+m_k^J(i)a_{ij}^\top
\nonumber\\
&\hspace{20mm}
+a_{ij}m_k^J(i)^\top
+w_k(i)a_{ij}a_{ij}^\top
\big\},
\label{eq:supp_JJ_recursion}\\
M_{k+1}^{GG}(j)
={}&
\sum_iR_{ij}
\big\{
M_k^{GG}(i)
+m_k^G(i)g_j^\top
\nonumber\\
&\hspace{20mm}
+g_jm_k^G(i)^\top
+w_k(i)g_jg_j^\top
\big\}.
\label{eq:supp_GG_recursion}
\end{align}

Finally, because $D_k$ is itself an additive matrix
reward,
\[
D_{k+1}=D_k+g_jg_j^\top,
\]
its unnormalized endpoint moment satisfies
\begin{equation}
M_{k+1}^D(j)
=
\sum_iR_{ij}
\{M_k^D(i)+w_k(i)g_jg_j^\top\}.
\label{eq:supp_D_recursion}
\end{equation}

Equations~\eqref{eq:supp_probability_recursion},
\eqref{eq:supp_jump_first_recursion},
\eqref{eq:supp_occupation_first_recursion},
\eqref{eq:supp_JG_recursion},
\eqref{eq:supp_JJ_recursion},
\eqref{eq:supp_GG_recursion}, and
\eqref{eq:supp_D_recursion}
form a closed forward system for the endpoint
probabilities and the required first and second
reward moments.

\paragraph{Initialization and sequential propagation.}

At $k=0$, the embedded path consists only of the
observed starting state $a$. Hence
\[
J_0=0,\qquad
G_0=g(a),\qquad
D_0=g(a)g(a)^\top.
\]

The initial endpoint-specific quantities are therefore
\begin{align}
w_0(a)&=1,
&m_0^J(a)&=0,
&m_0^G(a)&=g(a),\\
M_0^{JJ}(a)&=0,
&M_0^{JG}(a)&=0,
&M_0^{GG}(a)=M_0^D(a)&=g(a)g(a)^\top,
\end{align}
with all quantities initialized to zero at other
endpoint states.

Starting from these values, all reward moments are
updated jointly from step $k$ to step $k+1$.
The resulting moments at each step can be accumulated
immediately with the corresponding Poisson probability.
The recursion is therefore not restarted at the initial
state for each candidate value of $k$.

\subsubsection{Conditional score moments and observed information}

Let $b$ denote the observed terminal state, and let
\[
p_k
=
P\{\operatorname{Poisson}(\lambda t_h)=k\}
=
e^{-\lambda t_h}
\frac{(\lambda t_h)^k}{k!}.
\]

Under exact uniformization, the transition probability
between the observed endpoint states is
\[
P_{ab}(t_h)
=
\sum_{k=0}^{\infty}p_kw_k(b).
\]

For numerical evaluation, truncate the Poisson series
at $K$ and define
\begin{equation}
d_{h,K}
=
\sum_{k=0}^Kp_kw_k(b).
\label{eq:supp_truncated_transition_probability}
\end{equation}
This is the truncated approximation to $P_{ab}(t_h)$.

The corresponding normalized truncated bridge weights are
\begin{equation}
P_K(K=k\mid\mathcal O_h)
=
\frac{p_kw_k(b)}{d_{h,K}},
\qquad k=0,\ldots,K,
\label{eq:supp_truncated_bridge_weights}
\end{equation}
provided $d_{h,K}>0$.

Because the propagated moments already include the
embedded-path probabilities, they are multiplied by
$p_k/d_{h,K}$ when accumulated across the possible
Poisson event counts.

Using Equation~\eqref{eq:supp_score_first_path},
the truncated conditional first score moment is
\begin{equation}
\begin{aligned}
\mu_{h,K}
&:=
E_K(S_{c,h}\mid\mathcal O_h)\\
&=
\sum_{k=0}^K
\frac{p_k}{d_{h,K}}
\{m_k^J(b)+c_km_k^G(b)\}.
\end{aligned}
\label{eq:supp_bridge_score_first}
\end{equation}

Similarly, combining
Equations~\eqref{eq:supp_score_second_path}
and~\eqref{eq:supp_score_product_expansion}
gives the truncated conditional second score moment
\begin{align}
B_{h,K}
&:=
E_K(S_{c,h}S_{c,h}^\top\mid\mathcal O_h)
\nonumber\\
&=
\sum_{k=0}^K\frac{p_k}{d_{h,K}}
\Big[
M_k^{JJ}(b)
+c_k\{M_k^{JG}(b)+M_k^{JG}(b)^\top\}
\nonumber\\
&\hspace{19mm}
+c_k^2M_k^{GG}(b)
+\frac{t_h^2}{L^2(L+1)}
\{LM_k^D(b)-M_k^{GG}(b)\}
\Big],
\qquad L=k+1.
\label{eq:supp_bridge_score_moments}
\end{align}

The expected complete-data Hessian depends only on the
conditional expected state occupation times, which are
already available from the E-step. Specifically, from
Equation~\eqref{eq:supp_complete_hessian}, the expected
Hessian block corresponding to transition $r\to s$ is
\begin{equation}
E\{H_{c,h}^{(rs,rs)}\mid\mathcal O_h\}
=
-q_{rs,h}x_hx_h^\top
E(T_{r,h}\mid\mathcal O_h),
\label{eq:supp_expected_hessian_block}
\end{equation}
where the conditional expected occupation time is
computed as described in Equation~\eqref{eq:supp_ET}.
The off-diagonal blocks of the expected complete-data
Hessian are zero.

For consistency with the truncated uniformization
calculations, let
\[
\overline H_{h,K}
=
E_K(H_{c,h}\mid\mathcal O_h)
\]
denote the expected complete-data Hessian evaluated
using the same truncation level $K$ as the E-step.
Its transition-specific diagonal blocks are
\begin{equation}
\overline H_{h,K}^{(rs,rs)}
=
-q_{rs,h}x_hx_h^\top
E_K(T_{r,h}\mid\mathcal O_h).
\label{eq:supp_truncated_expected_hessian}
\end{equation}
Thus, the expected Hessian is obtained directly from
the conditional expected occupation times, without
additional moment recursions.

Combining the expected Hessian with the conditional
score moments $\mu_{h,K}$ and $B_{h,K}$ derived above,
the interval-specific Louis-type information
approximation is
\begin{equation}
I_{\mathrm{Louis},h,K}(\theta)
=
-\overline H_{h,K}
-
\left\{
B_{h,K}
-
\mu_{h,K}\mu_{h,K}^\top
\right\}.
\label{eq:supp_truncated_louis}
\end{equation}

Summing over all observation intervals gives
\begin{equation}
I_{\mathrm{Louis},K}(\theta)
=
\sum_h I_{\mathrm{Louis},h,K}(\theta).
\label{eq:supp_total_truncated_information}
\end{equation}

As the truncation level increases, the resulting
information approximation approaches the exact
observed information under appropriate convergence
conditions. The truncation level is selected to
control the uniformization approximation error.

The estimated covariance matrix and corresponding
Wald standard errors are computed as
\begin{equation}
\widehat{\operatorname{Var}}(\hat\theta)
=
I_{\mathrm{Louis},K}(\hat\theta)^{-1},
\qquad
SE(\hat\theta_l)
=
\sqrt{
\left[
I_{\mathrm{Louis},K}(\hat\theta)^{-1}
\right]_{ll}
}.
\label{eq:supp_louis_standard_errors}
\end{equation}
These expressions assume that the estimated
information matrix is nonsingular.

\clearpage
\section{Patient-Level CAV Pathway Probabilities}

For each of the 148 consecutive observed intervals with endpoints in states 1
and 4, we used the fitted coefficients and covariates measured at the
beginning of the interval to calculate exact posterior probabilities for the
collapsed clinical paths $1\to4$, $1\to2\to4$, and
$1\to2\to3\to4$. The exact dynamic program sums over all uniformization
sequences whose consecutive distinct states equal the specified clinical
path. Because return transitions are permitted, the three selected paths are
not exhaustive. We therefore report
\[
p_{\mathrm{other}}
=1-p_{14}-p_{124}-p_{1234},
\]
which contains all other, including recurrent, clinical paths.

\footnotesize
\setlength{\LTcapwidth}{\textwidth}
\begin{longtable}{rrrrr}
\caption{Exact posterior probabilities of selected latent clinical pathways for consecutive observed intervals with endpoints in states 1 and 4. The other/recurrent probability is one minus the sum of the three selected-path probabilities.}\label{tab:supp_cav_pathways}\\
\toprule
Patient ID & $p_{14}$ & $p_{124}$ & $p_{1234}$ & $p_{\mathrm{other}}$ \\
\midrule
\endfirsthead
\multicolumn{5}{c}{\tablename\ \thetable\ continued}\\
\toprule
Patient ID & $p_{14}$ & $p_{124}$ & $p_{1234}$ & $p_{\mathrm{other}}$ \\
\midrule
\endhead
\midrule\multicolumn{5}{r}{Continued on next page}\\
\endfoot
\bottomrule
\endlastfoot
100020 & 0.4972 & 0.3262 & 0.1715 & 0.0050 \\
100021 & 0.9722 & 0.0263 & 0.0015 & 0.0000 \\
100029 & 0.9667 & 0.0259 & 0.0067 & 0.0007 \\
100036 & 0.9018 & 0.0725 & 0.0212 & 0.0045 \\
100038 & 0.9338 & 0.0479 & 0.0148 & 0.0035 \\
100040 & 0.9767 & 0.0179 & 0.0032 & 0.0022 \\
100046 & 0.8979 & 0.0668 & 0.0307 & 0.0047 \\
100051 & 0.8526 & 0.0852 & 0.0543 & 0.0079 \\
100052 & 0.8209 & 0.0654 & 0.0696 & 0.0441 \\
100053 & 0.8265 & 0.0813 & 0.0832 & 0.0090 \\
100056 & 0.4951 & 0.3165 & 0.1743 & 0.0141 \\
100064 & 0.6628 & 0.1044 & 0.1485 & 0.0843 \\
100066 & 0.6361 & 0.1052 & 0.1622 & 0.0966 \\
100067 & 0.7530 & 0.0980 & 0.1328 & 0.0163 \\
100068 & 0.6651 & 0.0973 & 0.1807 & 0.0570 \\
100070 & 0.6242 & 0.2733 & 0.0967 & 0.0058 \\
100074 & 0.6160 & 0.1007 & 0.1643 & 0.1190 \\
100076 & 0.8959 & 0.0850 & 0.0171 & 0.0021 \\
100082 & 0.8797 & 0.0723 & 0.0384 & 0.0097 \\
100087 & 0.9790 & 0.0184 & 0.0020 & 0.0006 \\
100088 & 0.6989 & 0.2053 & 0.0842 & 0.0116 \\
100096 & 0.5199 & 0.0815 & 0.2138 & 0.1849 \\
100101 & 0.7889 & 0.1487 & 0.0604 & 0.0019 \\
100106 & 0.8467 & 0.0867 & 0.0573 & 0.0093 \\
100107 & 0.9140 & 0.0438 & 0.0242 & 0.0180 \\
100111 & 0.9746 & 0.0221 & 0.0028 & 0.0005 \\
100115 & 0.8097 & 0.1150 & 0.0639 & 0.0114 \\
100119 & 0.9291 & 0.0441 & 0.0217 & 0.0051 \\
100150 & 0.8563 & 0.1105 & 0.0307 & 0.0025 \\
100163 & 0.6787 & 0.1094 & 0.1479 & 0.0640 \\
100175 & 0.5656 & 0.1866 & 0.2288 & 0.0190 \\
100180 & 0.8808 & 0.0686 & 0.0357 & 0.0149 \\
100184 & 0.7140 & 0.1825 & 0.0708 & 0.0327 \\
100186 & 0.7213 & 0.1216 & 0.0998 & 0.0572 \\
100187 & 0.9706 & 0.0212 & 0.0046 & 0.0035 \\
100189 & 0.9756 & 0.0198 & 0.0039 & 0.0007 \\
100202 & 0.9424 & 0.0482 & 0.0085 & 0.0009 \\
100205 & 0.5135 & 0.1535 & 0.2841 & 0.0490 \\
100215 & 0.6940 & 0.0570 & 0.0940 & 0.1550 \\
100224 & 0.9527 & 0.0402 & 0.0059 & 0.0012 \\
100229 & 0.8740 & 0.0627 & 0.0501 & 0.0131 \\
100239 & 0.9375 & 0.0344 & 0.0140 & 0.0140 \\
100240 & 0.9212 & 0.0516 & 0.0183 & 0.0089 \\
100241 & 0.8572 & 0.0781 & 0.0483 & 0.0164 \\
100246 & 0.3165 & 0.2711 & 0.2857 & 0.1267 \\
100247 & 0.5403 & 0.1255 & 0.2694 & 0.0647 \\
100249 & 0.8411 & 0.0735 & 0.0486 & 0.0369 \\
100254 & 0.9321 & 0.0587 & 0.0077 & 0.0016 \\
100257 & 0.5765 & 0.1714 & 0.2183 & 0.0339 \\
100261 & 0.7011 & 0.1531 & 0.0935 & 0.0524 \\
100263 & 0.1915 & 0.3859 & 0.3243 & 0.0983 \\
100270 & 0.7186 & 0.2433 & 0.0377 & 0.0004 \\
100276 & 0.3865 & 0.3703 & 0.2043 & 0.0389 \\
100278 & 0.9105 & 0.0671 & 0.0180 & 0.0044 \\
100281 & 0.8199 & 0.0477 & 0.0450 & 0.0873 \\
100293 & 0.9477 & 0.0429 & 0.0068 & 0.0026 \\
100296 & 0.8451 & 0.0738 & 0.0479 & 0.0331 \\
100298 & 0.4550 & 0.1116 & 0.2040 & 0.2294 \\
100300 & 0.9775 & 0.0182 & 0.0033 & 0.0011 \\
100305 & 0.4345 & 0.4762 & 0.0880 & 0.0012 \\
100307 & 0.3823 & 0.3112 & 0.1834 & 0.1230 \\
100311 & 0.2936 & 0.3491 & 0.3379 & 0.0194 \\
100316 & 0.7493 & 0.2164 & 0.0289 & 0.0053 \\
100327 & 0.9090 & 0.0411 & 0.0227 & 0.0272 \\
100330 & 0.1669 & 0.2900 & 0.4678 & 0.0754 \\
100332 & 0.2462 & 0.2821 & 0.3680 & 0.1037 \\
100338 & 0.6537 & 0.1510 & 0.1093 & 0.0861 \\
100341 & 0.8473 & 0.1028 & 0.0459 & 0.0040 \\
100342 & 0.7741 & 0.1602 & 0.0469 & 0.0188 \\
100359 & 0.7769 & 0.1587 & 0.0458 & 0.0185 \\
100363 & 0.9863 & 0.0128 & 0.0006 & 0.0003 \\
100364 & 0.7891 & 0.1382 & 0.0574 & 0.0153 \\
100365 & 0.4859 & 0.1530 & 0.2586 & 0.1025 \\
100366 & 0.7460 & 0.1677 & 0.0561 & 0.0301 \\
100370 & 0.8907 & 0.0594 & 0.0275 & 0.0224 \\
100384 & 0.9030 & 0.0652 & 0.0286 & 0.0032 \\
100386 & 0.7137 & 0.1017 & 0.1222 & 0.0624 \\
100397 & 0.3938 & 0.2602 & 0.2147 & 0.1314 \\
100399 & 0.9099 & 0.0466 & 0.0272 & 0.0164 \\
100418 & 0.8594 & 0.1157 & 0.0201 & 0.0048 \\
100419 & 0.5335 & 0.2739 & 0.1393 & 0.0533 \\
100425 & 0.5349 & 0.1109 & 0.2637 & 0.0904 \\
100438 & 0.6744 & 0.1510 & 0.1460 & 0.0286 \\
100445 & 0.3284 & 0.2068 & 0.3727 & 0.0922 \\
100465 & 0.1407 & 0.5019 & 0.2719 & 0.0855 \\
100469 & 0.7716 & 0.1237 & 0.0828 & 0.0218 \\
100498 & 0.9389 & 0.0446 & 0.0128 & 0.0037 \\
100513 & 0.9812 & 0.0145 & 0.0021 & 0.0022 \\
100533 & 0.8618 & 0.0594 & 0.0505 & 0.0282 \\
100549 & 0.9996 & 0.0004 & 0.0000 & 0.0000 \\
100550 & 0.9960 & 0.0038 & 0.0001 & 0.0001 \\
100557 & 0.4024 & 0.2295 & 0.3265 & 0.0416 \\
100580 & 0.9990 & 0.0010 & 0.0000 & 0.0000 \\
100587 & 0.9479 & 0.0403 & 0.0101 & 0.0017 \\
100591 & 0.9315 & 0.0432 & 0.0206 & 0.0047 \\
100600 & 0.8857 & 0.0982 & 0.0137 & 0.0024 \\
100607 & 0.9966 & 0.0033 & 0.0001 & 0.0000 \\
100610 & 0.7291 & 0.0663 & 0.1010 & 0.1035 \\
100613 & 0.9938 & 0.0058 & 0.0003 & 0.0001 \\
100621 & 0.5868 & 0.1739 & 0.2124 & 0.0269 \\
100626 & 0.7729 & 0.0705 & 0.0928 & 0.0638 \\
100627 & 0.9970 & 0.0029 & 0.0001 & 0.0001 \\
100632 & 0.8732 & 0.0916 & 0.0341 & 0.0011 \\
100636 & 0.9564 & 0.0275 & 0.0083 & 0.0079 \\
100637 & 0.8569 & 0.0675 & 0.0602 & 0.0154 \\
100638 & 0.8770 & 0.0620 & 0.0486 & 0.0124 \\
100642 & 0.9686 & 0.0240 & 0.0058 & 0.0016 \\
100644 & 0.9331 & 0.0476 & 0.0148 & 0.0044 \\
100646 & 0.9721 & 0.0220 & 0.0048 & 0.0011 \\
100649 & 0.7842 & 0.0717 & 0.0907 & 0.0533 \\
100650 & 0.9970 & 0.0028 & 0.0001 & 0.0001 \\
100657 & 0.9996 & 0.0004 & 0.0000 & 0.0000 \\
100662 & 0.9959 & 0.0040 & 0.0001 & 0.0000 \\
100663 & 0.9942 & 0.0055 & 0.0003 & 0.0001 \\
100704 & 0.9901 & 0.0091 & 0.0008 & 0.0001 \\
100706 & 0.9786 & 0.0164 & 0.0027 & 0.0023 \\
100708 & 0.9947 & 0.0050 & 0.0002 & 0.0000 \\
100725 & 0.9782 & 0.0174 & 0.0030 & 0.0015 \\
100731 & 0.5511 & 0.1186 & 0.2619 & 0.0684 \\
100737 & 0.8873 & 0.0560 & 0.0404 & 0.0164 \\
100742 & 0.9379 & 0.0401 & 0.0176 & 0.0044 \\
100747 & 0.9878 & 0.0105 & 0.0010 & 0.0006 \\
100752 & 0.7898 & 0.0671 & 0.0820 & 0.0611 \\
100753 & 0.9975 & 0.0024 & 0.0001 & 0.0000 \\
100757 & 0.9996 & 0.0004 & 0.0000 & 0.0000 \\
100760 & 0.9382 & 0.0405 & 0.0177 & 0.0036 \\
100767 & 0.9997 & 0.0003 & 0.0000 & 0.0000 \\
100768 & 0.9995 & 0.0005 & 0.0000 & 0.0000 \\
100772 & 0.9815 & 0.0160 & 0.0025 & 0.0001 \\
100778 & 0.8935 & 0.0579 & 0.0401 & 0.0085 \\
100793 & 0.9998 & 0.0002 & 0.0000 & 0.0000 \\
100794 & 0.8603 & 0.0663 & 0.0578 & 0.0156 \\
100796 & 0.9896 & 0.0094 & 0.0008 & 0.0002 \\
100801 & 0.9643 & 0.0256 & 0.0068 & 0.0033 \\
100806 & 0.9890 & 0.0099 & 0.0009 & 0.0002 \\
100812 & 0.9828 & 0.0142 & 0.0020 & 0.0010 \\
100819 & 0.9955 & 0.0042 & 0.0002 & 0.0001 \\
100823 & 0.9977 & 0.0022 & 0.0000 & 0.0000 \\
100830 & 0.9957 & 0.0041 & 0.0002 & 0.0001 \\
100832 & 0.8975 & 0.0603 & 0.0398 & 0.0024 \\
100837 & 0.9957 & 0.0042 & 0.0002 & 0.0000 \\
100848 & 0.9645 & 0.0243 & 0.0062 & 0.0051 \\
100854 & 0.9929 & 0.0064 & 0.0004 & 0.0003 \\
100858 & 0.9990 & 0.0010 & 0.0000 & 0.0000 \\
100860 & 0.9964 & 0.0035 & 0.0001 & 0.0000 \\
100880 & 0.9978 & 0.0021 & 0.0000 & 0.0000 \\
100892 & 0.9990 & 0.0010 & 0.0000 & 0.0000 \\
100897 & 0.9983 & 0.0016 & 0.0000 & 0.0000 \\
\end{longtable}

\normalsize

\bibliographystyle{plainnat}
\bibliography{references}